\documentclass[onecolumn,secnumarabic,amssymb,amsmath,amsfonts nobibnotes, aps, prd,preprintnumbers]{revtex4-1}

\usepackage{epsfig,graphicx}
\usepackage{chngcntr}
\usepackage{rotating}
\usepackage{wrapfig}
\usepackage{parskip} 
\usepackage{graphicx}
\usepackage{caption}
\usepackage{subcaption}
\usepackage[linkcolor=blue, citecolor=red, urlcolor=blue]{hyperref}

\begin{document}

\title{Vacuum stability  in the Standard Model with vector-like fermions}%

\author{\sf Ash Arsenault$^{1,2}$\footnote{ashley.n.arsenault@gmail.com},\; Kivanc Y. Cingiloglu$^{1,}$\footnote{kivanc.cingiloglu@concordia.ca},\; Mariana Frank$^{1,}$\footnote{mariana.frank@concordia.ca}}%

\affiliation{$^1$ Department of Physics,  
			Concordia University, 7141 Sherbrooke St.West,\\
			Montreal, Quebec, Canada H4B 1R6.\\}
\affiliation{$^2$ Veterans Affairs Canada P.O.Box 7700 Charlottetown, \\ PE, Canada C1A 8M9}
\date{\today}%
			\begin{abstract}
The discovery of Standard-Model like Higgs at 125 GeV may raise more questions than the answers it provides. In particular, the hierarchy problem remains unsolved, and the Standard Model Higgs quartic self-coupling becomes negative below the
Planck scale, necessitating new physics beyond the Standard Model. In this work we investigate a popular scenario, extensions of the Standard Model with vector-like fermion fields, such as the ones present in models with extra dimensions or in Higgs composite models, using a model independent approach. Since fermions decrease the Higgs quartic coupling at high energies, only exacerbating
the self-coupling problem, we introduce first an additional scalar, which by itself
is enough to overcome the vacuum stability limit, and then explore the effects of vector-like fermions in singlet, doublet and triplet representations. 
 For each model, we  identify the allowed fermion masses and  mixing angles with the third family fermions  
required to satisfy the vacuum stability condition, and compare different representations.  Allowed fermion masses emerge at around 1 TeV, raising hope that these will be found at the LHC. We also examine corrections to oblique parameters $\mathbb{S}$ and $\mathbb{T}$ from additional scalar and vector-like quarks which also impose constraints on mixing and mass splitting of both sectors., but these restrictions are relatively weak compared to the vacuum stability.
\end{abstract}

\maketitle

\bigskip


\section{Introduction}
 \label{sec:intro}

Ever since the Higgs boson was discovered  at the CERN Large
Hadron Collider (LHC), confirming at last the last remaining puzzle  of the Standard Model (SM) \cite{cms-atlas},  the observed
mass of the Higgs boson combined with the mass of the top quark, $m_t$, have caused concern because, as the theory stands,  it violates stability of the electroweak vacuum  \cite{stability}. In the SM there is a single Higgs $h$ with effective potential is characterized by two parameters only, the Higgs (mass)$^2$, $\kappa^2$ and its self coupling $\lambda$, $V=\kappa^2 h^2 +\lambda h^4$. The self-coupling $\lambda$ can become negative at larger scales, so the potential becomes unbounded
from below, and there is no resulting stability. Theoretical considerations indicate that if the validity of the SM
is extended to $M_{\rm {Planck}}$, a second, deeper minimum is located near the Planck scale
such that the electroweak vacuum is metastable, i.e., the transition lifetime of the
electroweak vacuum to the deeper minimum is finite with lifetime $\sim 10^{300}$ years \cite{stability}. \\ If the electroweak vacuum is metastable then
Higgs cannot play the role of inflaton.   Explanations involving a long lived-universe, where vacuum instability is not important, were proved to be faulty. Without vacuum
stability, fluctuations in the Higgs field during inflation and in the hot early
universe would have taken most of the universe into an anti-De Sitter phase,
yielding a massive collapse, and the expansion of the universe would never have
occurred.
The result of this is that either the SM must be incorrect or flawed in some way, or at the very
least, that new physics beyond the SM which 
alters the Higgs potential so that it enhances its stability must exist at higher energies. Thus extra degrees of freedom are needed
for the SM to explain the inflation of the Universe.

Minimal extensions of the SM which stabilize the Higgs vacuum are the most common
theories which attempt to solve the Higgs mass problem. The correlation between the Higgs mass
and vacuum stability is highly dependent on bosonic interactions. For instance, a  model \cite{Gunion:2002zf} with two
Higgs doublets and large soft Higgs mass terms, satisfying the electroweak symmetry breaking conditions, has a stable vacuum and 
 decay branching ratios that are very close to the SM ones, and this only one example. 

The question remains if models with additional fermions, present in most beyond the SM  scenarios, can survive stability constraints, and if so, what are the restrictions imposed on their masses and mixing (if any) with the SM particles. 

To investigate how  the hierarchy problem may be fixed, and what are the implications for vacuum stability, one could proceed by assuming a theory which supersedes the SM emerging at higher energies, such a supersymmetry. (Note however that minimal supersymmetry has its own difficulties with accommodating a Higgs boson of mass 125 GeV.) Or one could study the effect of adding particles to the SM, coupled in the simplest way, and investigate the conditions on their masses and couplings as emerging from vacuum stability conditions, as  a simple and elegant way to obtain information about new particles and interactions without assuming complicated frameworks.

The latter is the approach  we wish to follow in this article, and we investigate inclusion of one additional generation of vector-like fermions, {\it i.e.}, fermions whose left-handed and right-handed components transform the same way under $SU(3)_c \times SU(2)_L \times U(1)_Y$. Unlike sequential fourth-generation quarks, which are ruled out by  the one-loop induced Higgs production and decay mechanisms (the
gluon fusion production and diphoton decay of the Higgs) \cite{AguilarSaavedra:2002kr}, indirect bounds on vector-like quarks are
much weaker. In particular, vector-like fermions can acquire a large Dirac mass without introducing a large
Yukawa coupling to the Higgs.

Vector-like fermions appear in the context of many models of New Physics \cite{Ellis:2014dza}. In warped or universal extra dimensional models, vector-like fermions appear as KK excitations of bulk fields \cite{Gopalakrishna:2013hua}, in Composite Higgs models,  vector-like quarks emerge as excited resonances of bound states that form SM particles \cite{Contino:2006qr, Anastasiou:2009rv}, in Little Higgs Models,  they are partners of the ordinary fermions within larger group representations and charged under the group \cite{ArkaniHamed:2002qy}, and in non-minimal supersymmetric extensions of the SM, they can increase the Higgs mass through loop corrections without adversely affecting electroweak precision \cite{Martin:2009bg}.  Vector-like coloured particles are consistent with perturbative
gauge coupling unification and are often invoked to explain discrepancies in the data, such as 
the  $t{\bar t}H$ anomaly \cite{Couture:2017mbd}.

Vector-like particles have been considered before in the context of stabilizing the vacuum of the SM in \cite{Xiao:2014kba}, and to help explain the observed excess at 750 GeV \cite{Dhuria:2015ufo,Zhang:2015uuo}. However, only particular representations have been considered \cite{Carmi:2012yp}, and a complete interplay of all possible vector-like quark representations and the SM does not exit at present. We redress this here, and analyze the restrictions on the masses and mixing angles for the all anomaly-free representations of vector-like quarks, as well as the associated  boson field which is added to the SM for vacuum stabilization. In addition, we test the effects and restrictions induced by the vector-like fermions on the electroweak precision observables, $\mathbb{S, T}$ and $\mathbb{U}$.

Our work is organized as follows. In Sec. \ref{sec:VSSM} we outline briefly the vacuum stability problem in the SM,  and in \ref{subsec:VSSM+H}  its resolution with an additional singlet scalar. In Sec. \ref{sec:VSSM+F} we then introduce all anomaly-free vector-like quark representations,  their interaction Lagrangians, and derive their masses and mixing angles (assumed to be with the third generation quarks only). We then proceed to analyze  the effects on vacuum stability of introducing singlet, doublet and triplet representations, respectively. In Sec. \ref{sec:electroweakprecision} we give the expressions and analyze the effects of the additional fields on the electroweak precision observables.  We conclude in Sec. \ref{sec:conclusion}, and leave the expressions for the relevant RGEs for the models  studied to the Appendix (sec \ref{sec:appendix}).

\section{Vacuum Stability in the SM}
\label{sec:VSSM}
\subsection{The Higgs potential}
\label{subsec:SMHiggs}
In the SM, interactions with the Higgs field are specified by the Higgs potential, which has the form
\begin{equation}
V(\phi)=-\frac12 \kappa^2 \Phi^2 +\frac14 \lambda \Phi^4 +\Delta V,
\end{equation}
with $\kappa$ and $\lambda$ the quadratic and quartic couplings,   $\Delta V$ the tree-level correction terms, and $\Phi$ the Higgs field given by
\begin{equation}
\Phi=\left(\begin{array}{c} \Phi^+ \\ \Phi^0 \end{array} \right) =\left(\begin{array}{c} \Phi^+ \\ (v+h^0+iG^0)/\sqrt{2} \end{array} \right)
 \label{eq:sm_higgs}
\end{equation}
where $v=246$ GeV is the Higgs vacuum expectation value (VEV). Vacuum stability requires $V^\prime(\phi)>0$, or equivalently, $\lambda(\mu)>0$, with $\lambda$ the running Higgs coupling which depends on the scale, $\mu$, at which the SM breaks down.
The issue is that the Higgs quadratic self-coupling is renormalized not only by itself ($\lambda$ increasing as the energy scale increases), but also by the Higgs (Yukawa) coupling to the top quark, which tends to drive it to smaller, even negative values at high scales $\mu$.  At leading orders \cite{Altarelli:1994rb}
\begin{equation}
\lambda(\mu)
\simeq \lambda(\mu_0)-\frac{3m_t^2}{2 \pi v^4} \log \left(\frac{\mu}{\mu_0}\right)\, , 
\end{equation}
yielding an estimate for the energy scale where $\lambda$ will become negative \cite{Altarelli:1994rb}
\begin{equation}
\log \left(\frac{\mu}{\mu_0}\right)=9.4+0.7 (m_H-125.15)-1.0 (m_t-173.34)+0.3 \left(\frac{\alpha_s(m_Z)-0.1184}{0.0007}\right)
\end{equation}
with energy scales and masses measured in GeV. Taking into account all the uncertainties in the measurements of $m_H, m_t$ and $\alpha_s$, the scale at which the SM fails is $\mu =10^{(9.4 \pm 1.1)} $ GeV. Possible remedies include higher-dimensional operators in the effective theory, or additional symmetries (such as supersymmetry, which is associated to a new scale), or addition of new particles to the SM.

More explicitly, the variation of the coupling $\lambda$ with the energy scale involves evaluation of the one-loop beta function describing the running of $\lambda$ is \cite{Altarelli:1994rb}
\begin{equation}
\frac{d \lambda(\mu)}{d \ln \mu}=\frac{1}{16 \pi^2}\left[4\lambda^2+12 \lambda y_t^2-36 y_t^2-9 \lambda g_1^2-3 \lambda g_2^2 +\frac94g_2^2 +\frac92g_1^2 g_2^2 +\frac{27}{4}g_1^4 \right],
\end{equation}
with $g_1, g_2, g_3$ the coupling constants for $U(1)_Y, SU(2)_L, SU(3)_c$ and $y_t$ the Yukawa coupling of the top quark. The renormalization group equations for these parameters are \cite{Tang:2013bz}:
\begin{eqnarray}
\label{eq:gauge_rge}
\frac{dg_i(\mu)}{d \ln \mu}=\frac{1}{16 \pi^2}b_ig_i^3,~b=(41/6, -19/6,-7)\\
\frac{dy_t(\mu)}{d \ln \mu}=\frac{y_t}{16 \pi^2}\left[\frac92 y_t^2-\frac94 g_2^2-\frac{17}{12}g_1^2-8g_3^2\right],
\end{eqnarray}
with initial conditions
\begin{eqnarray}
g_1^2(\mu_0)&=&4 \pi \alpha, \qquad g_2^2(\mu_0)=4 \pi \alpha\left( \frac{1}{\sin \theta_W}+1 \right),  \qquad g_3^2(\mu_0)=4 \pi \alpha_s 
\nonumber \\
y_t(\mu_0)&=&\frac{\sqrt{2}m_t}{v}, \qquad \lambda (\mu_0)=\frac{3m_H^2}{v^2} [1+\delta_\lambda(\mu_0)].
\end{eqnarray}
Here $\alpha, \alpha_s$ are the weak and strong coupling constants, $\sin \theta_W=0.2312$ is the Weinberg angle,  and we set $\mu_0=m_Z=91.188$ GeV. The radiative decay constant is \cite{Sirlin:1985ux}
\begin{equation}
\delta_\lambda(\mu)= \frac{G_F m_Z^2}{8 \sqrt{2} \pi^2} \left[\xi f_1(\xi, \mu)+f_0(\xi, \mu) +\xi^{-1} f_{-1}(\xi, \mu)\right],
\end{equation}
with $\xi= m_H^2/m_Z^2$, $G_F=1.16635 \times 10^{-5}$ GeV$^{-2}$ and
\begin{eqnarray}
f_1(\xi, \mu)&=&6 \ln \frac{\mu^2}{m_H^2}+\frac32 \ln \xi-\frac12 Z\left(\frac{1}{\xi}\right)- Z\left(\frac{c^2}{\xi}\right)-\ln c^2+\frac92 \left(\frac{25}{9}-
\sqrt{\frac13} \pi \right)\, \nonumber \\
f_0(\xi, \mu)&=&6 \ln \frac{\mu^2}{m_Z^2}\left [1+2c^2-2 \frac{m_t^2}{m_Z^2} \right]+\frac{3c^2 \xi}{\xi-c^2}+2Z\left(\frac{1}{\xi}\right) +4c^2 Z\left(\frac{c^2}{\xi}\right) +\frac{3c^2 \ln c^2}{s^2} \nonumber \\
&+&12c^2 \ln c^2-\frac{15}{2}(1+2c^2) -3 \frac{m_t^2}{m_Z^2} \left [ 2 Z\left(\frac{m_t^2}{\xi m_Z^2}\right)+4 \ln \frac{m_t^2}{m_Z^2} -5 \right], \nonumber \\
f_{-1}(\xi, \mu)&=&6 \ln \frac{\mu^2}{m_Z^2}\left [1+2c^4-24\frac{m_t^2}{m_Z^2} \right] -6Z\left(\frac{1}{\xi}\right)- 12 c^4 Z\left(\frac{c^2}{\xi}\right)-12 c^4 \ln c^2  \nonumber \\
&+&8(1+2c^4) + \left [  Z\left(\frac{m_t^2}{\xi m_Z^2}\right)+ \ln \frac{m_t^2}{m_Z^2} -2 \right]\,
\end{eqnarray}
with 
\begin{eqnarray}
Z(z)&=&\left \{ \begin{array}{lr} 2A\tan^{-1}(1/A), &z>1/4\\
A \ln [(1+A)/(1-A)], &z<1/4,
\end{array}\right.  \\
A&=&|1-4z|^1/2
\end{eqnarray} 
where $c, s$ are abbreviations for $\cos \theta_W,\, \sin \theta_W$.

Examining the running of the coupling parameters in this mode shows that  the scalar couplings
increase with increasing energy scales, while the Higgs coupling $\lambda$ 
decreases  and becomes negative at around $10^{10}$ GeV.

\subsection{Introducing  an additional boson}
\label{subsec:VSSM+H}
In this section, we consider the simplest remedy to the stability problem by extending the particle content of the   SM by an extra (singlet, as it is simplest)  scalar boson
 which interacts solely with the SM Higgs, and we
examine the constraints placed on its mass and mixing angle with the existing Higgs boson by the
Higgs vacuum stability condition. The addition of a boson provides a positive boost to the coupling parameter, counteracting the effect of the top quark and contributing towards repairing
the Higgs vacuum stability. 

In this scenario, the Higgs doublet $\Phi$ from Eq. \ref{eq:sm_higgs} interacts with the new scalar singlet $\chi$ 
\begin{equation}
\chi=\left ( u+\chi^0 \right)
\end{equation}
where $u$ is the singlet VEV, through
\begin{equation}
V(\phi, \chi)=-\kappa_H^2 \Phi^\dagger \Phi+\lambda_H( \Phi^\dagger \Phi )^2 -\frac{\kappa_S^2}{2} \chi^2 +\frac{\lambda_S}{4} \chi^4+\frac{\lambda_{SH}}{2} (\Phi^\dagger \Phi)^2 \chi^2.
\end{equation}
After symmetry breaking,  the singlet and doublet Higgs mix 
\begin{equation}
{\cal M}_{H,S}=  \left( \begin{array}{cc}2 \lambda_Hv^2& \lambda_{SH}vu \\ \lambda_H v u  &2 \lambda_S u^2 \end{array}
 \right)\, ,
 \end{equation}
yielding mass eigenvalues:
\begin{equation}
m^2_{H,S}=\lambda_Hv^2+\lambda_Su^2\mp\sqrt{(\lambda_Su^2-\lambda_Hv^2)^2+\lambda_{SH}^2u^2v^2}
\end{equation}
and the eigenvectors
\begin{equation}
\left(  \begin{array}{c} H \\ S  \end{array} \right )=  \left(  \begin{array}{cc} \cos \varphi & \sin \varphi \\ -\sin \varphi  & \cos \varphi \end{array}  \right)
\left( \begin{array}{c} \Phi\\ \chi  \end{array} \right )\, .
 \end{equation}

\begin{figure}
	\begin{center}
	\includegraphics[width=2.5in]{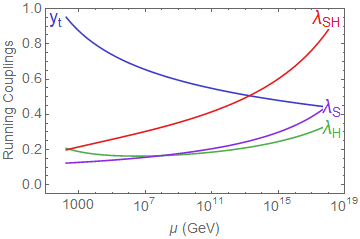}
	\end{center}
	\caption{The RGE running of the top Yukawa coupling and scalar couplings for the scalar boson model with $m_S=$ 1 TeV, $\sin\varphi=0.1$, $u=$ 2 TeV, and setting the starting point of the running  at $\mu_0=m_t$.}
	\label{fig:rgeS}
\end{figure}

As in the SM, we require $\lambda_H>0$ for a stable SM vacuum, and $\lambda_S>0$ for the new particle. In addition we impose that the potential is positive for asymptotically large values of the fields, 
\begin{equation}
\label{eq:singletstability}
\lambda_H>0, \qquad 0<\lambda_S<4 \pi, \qquad |\lambda_{SH}|>4 \pi.
\end{equation}
The Yukawa couplings can be expressed in terms of the physical masses as:
\begin{eqnarray}
\label{eq:1singletmasses}
\lambda_H&=&\frac{m_H^2 \cos^2 \varphi + m_S^2 \sin^2 \varphi}{2v^2} \, , \nonumber
 \\
\lambda_S&=&\frac{m_S^2 \cos^2 \varphi + m_H^2 \sin^2 \varphi}{2v^2}\, , \nonumber 
\\
\lambda_{SH}&=&\frac{m_S^2-m_H^2}{2uv} \sin 2 \varphi, 
\end{eqnarray}
with $m_H, \, m_S ~(v,u)$ the masses  (VEVs) of the physical fields, respectively and $\varphi$ their mixing angle.
Requiring perturbativity up to Planck scales, we apply the Yukawa and Higgs sector RGEs:
\begin{eqnarray}
\label{eq:rge1singlet}
\frac{dy_t^2}{d \ln \mu^2}&=& \frac{y_t^2}{16 \pi^2}\left (\frac{9y_t^2}{2}-\frac{17g_1^2}{20}-\frac{9g_2^2}{4}- 8g_3^2 \right)\, ,\nonumber\\
\frac{d \lambda_H}{d \ln \mu^2}&=& \frac{1}{16 \pi^2} \left[  \lambda_H \left (12 \lambda_H+6 y_t^2-\frac{9 g_1^2}{10}-\frac{9 g_2^2}{2} \right)+ 
\left( \frac{\lambda_{SH}^2}{4}-3 y_t^4 +\frac{ 27g_1^4}{400}+\frac{ 9g_2^4}{16} +\frac{ 9g_1^2 g_2^2 }{40} \right ) \right ] \, , \nonumber \\
 \frac{d \lambda_S}{d \ln \mu^2}&=&\frac{1}{16 \pi^2} \left (9 \lambda_S^2+\lambda_{SH}^2 \right) \, , \nonumber \\
 \frac{d \lambda_{SH}}{d \ln \mu^2}&=&\frac{\lambda_{SH}}{16 \pi^2} \left(2 \lambda_{SH}+6 \lambda_H +3 \lambda_S +3 \lambda_S +3 y_t^2 -\frac{9g_1^2}{20}- \frac{9g_1^2}{20} \right) ,
 \end{eqnarray}
Here $\lambda_H$ and $\lambda_S$ are the quartic self-couplings of $\Phi$ and $\chi$, and $\lambda_{SH}$ the coupling describing their mixing. Eqs. \ref{eq:1singletmasses} describe the coupling parameters at relatively small
energy scales, and therefore serve as initial conditions to these RGEs. 
\\Just as in the
SM, we ignore the contributions of all Yukawa couplings except for that of the top
quark, and also, we include electroweak radiative correction terms for increased accuracy. To this end, we
replace the top Yukawa coupling and Higgs self-coupling boundary conditions with \cite{Hempfling:1994ar}
\begin{eqnarray}
y_t&=&\frac{\sqrt{2} m_t}{v} [1+\Delta_t(\mu_0)]\, , \nonumber
\\
\lambda_H&=&\frac{m_H^2 \cos^2 \varphi + m_S^2 \sin^2 \varphi}{2v^2}[1+\Delta_H(\mu_0)]
\end{eqnarray}
where $\Delta_H (\mu)$ is the same correction as in the SM, and 
\begin{equation}
\Delta_t(\mu_0)= \Delta_W(\mu_0) +\Delta_{QED}(\mu_0) +\Delta_{QCD}(\mu_0) \, ,
\end{equation}
with 
\begin{eqnarray}
\Delta_W(\mu_0)&=&\frac{G_Fm_t^2}{16 \sqrt{2} \pi^2}\left (-9 \ln\frac{m_t^2}{\mu_0^2}-4 \pi \frac{m_H}{m_t}+11 \right )\, , \nonumber \\
\Delta_{QED}(\mu_0)&=&\frac{\alpha}{9 \pi} \left (3 \ln \frac{m_t^2}{\mu_0^2}-4 \right)\, , \nonumber \\
\Delta_{QCD}(\mu_0)&=&\frac{\alpha_s}{9 \pi} \left (3 \ln \frac{m_t^2}{\mu_0^2}-4 \right)\, , 
\end{eqnarray}
and we include the RGEs for the gauge couplings as in the SM, Eq. \ref{eq:gauge_rge}.
\\\\
Fig. \ref{fig:rgeS} illustrates the running of the coupling parameters for a typical set of parameter values.
Notice that in this model, the scalar couplings increase with increasing energy scales, compensating for the  SM  Higgs coupling, which  becomes negative at around $10^{10}$ GeV.  Therefore, the addition of an extra scalar boson to the SM rescues the theory from vacuum instability. 
\newpage

\begin{figure}[htbp]
	\centering
	\begin{subfigure}{.33\textwidth}\hspace{-1.5cm}
		\includegraphics[height=1.5in]{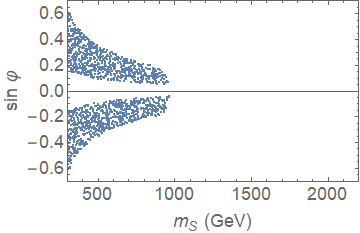}
		\caption{$u=$ 1 TeV.}
	\end{subfigure}\hspace{-0.8cm}
	\begin{subfigure}{.33\textwidth}
		\includegraphics[height=1.5in]{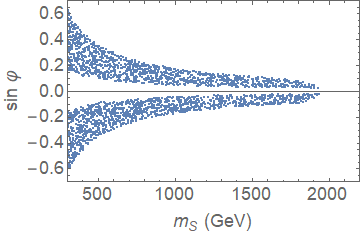}
		\caption{$u=$ 2 TeV.}
	\end{subfigure}\hspace{0.3cm}
	\begin{subfigure}{.33\textwidth}
		\includegraphics[height=1.5in]{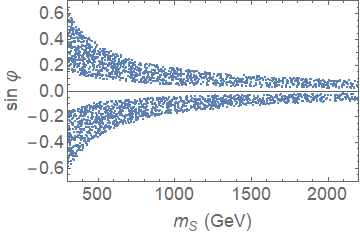}
		\caption{$u=$ 4 TeV.}
	\end{subfigure}
  \caption{ The allowed parameter space for the mass $m_S$ and mixing angle $\varphi$ with the SM Higgs for the additional scalar boson model, for different  vacuum
expectation values: $u=1$ TeV (left panel); $u=2$ TeV (middle panel); and $u=4$ TeV (right panel).}
  \label{fig:scalarBosonParams}
\end{figure}

Of course, we may investigate the mass and mixing angle of this singlet scalar with the Higgs boson by eliminating all parameter values that
do not satisfy Higgs vacuum stability.  For this we perform a scan over a broad parameter space
and disallow all parameter values which do not satisfy the vacuum stability conditions outlined in Eq. \ref{eq:singletstability}. The
resulting allowed parameter space is illustrated in Fig. \ref{fig:scalarBosonParams}.

Note that, while the mass region allowed for the additional boson for $u=1$ TeV is quite restricted, for larger VEVs it is quite large, and increasing with the new boson VEV. However, in all cases, the mixing with the SM Higgs boson is required to be non-zero ($\varphi \ne 0$), consistent with the observation of Higgs potential instability in the absence of the additional boson.
\section{Vacuum Stability in the SM with vector-like fermions}
\label{sec:VSSM+F}

\subsection{Theoretical Considerations}
\label{subsec:VLQtheory}
Using the stable potential in the previous section, we investigate the effect of introducing vector-like fermions  into this SM + additional scalar model.
Unlike SM-like (chiral) fermions which act as doublets under $SU(2)_L$ if left-handed and as singlets if
right-handed, and spoil the agreement of loop-induced production and decays of the SM Higgs, vector-like fermions have the same interactions regardless of chirality. They appear in many new physics models, such as models with extra dimensions, and many explanations put forward of deviations from SM physics include vector-like fermions.Thus it is important  that, when considering the addition of a vector-like fermions to the SM, the presence of a new scalar boson is essential to ensure the stability of the Higgs potential, otherwise, as the fermions decrease the effective self coupling,  the singular divergence of the Higgs quartic coupling would worsen compared to the one in the SM. As before we require the Higgs sector potential 
to be positive at asymptotically large values of the fields, up to Planck scale. The question we need to address is: what are the constraints on the masses of the vector-like fermions, and mixing angles with ordinary fermions, such as to maintain vacuum stability.
\newpage
The new states interact with the Higgs bosons through Yukawa interactions. The allowed anomaly-free multiplet states for the  vector-like quarks, together with their nomenclature, are listed in Table \ref{tab:VQrepresentations} \cite{Carmi:2012yp,Ellis:2014dza,Aguilar-Saavedra:2013qpa}. The first two representations are $U$-like and $D$-like singlets, the next three are doublets (one SM-like, two non-SM like), and the last two are triplets. The various representations are distinguished by their $SU(2)_L$ and hypercharge numbers.
\begin{table}[htbp]
\caption{\label{tab:VQrepresentations}\sl\small Representations of Vector-Like Quarks, with quantum numbers under $SU(2)_L \times U(1)_Y$.}
  \begin{center}
 \small
 \begin{tabular*}{0.99\textwidth}{@{\extracolsep{\fill}} c| ccccccc}
 \hline\hline
	Name &${\cal U}_1$ &${\cal D}_1$ &${\cal D}_2$ &${\cal D}_X$ &${\cal D}_Y$ &${\cal T}_X$ 
	&${\cal T}_Y$\\
  Type&Singlet &Singlet &Doublet&Doublet &Doublet &Triplet 
	&Triplet\\
	 \hline
	   &$T$ &$B$ &$\left ( \begin{array}{c} T \\ B \end{array} \right ) $ &$ \left (\begin{array}{c}  X \\ T  \end{array}\right)$ &$\left ( \begin{array}{c} B \\ Y\end{array} \right ) $ &$ \left (\begin{array}{c} X\\T \\ B \end{array} \right ) $
	  & $\left ( \begin{array}{c} T \\ B\\Y \end{array} \right )$\\
  \hline
  $SU(2)_L$ &1 &1 &2 &2 & 2 &3 &3 \\
  \hline
  $Y$ &$ 2/3$ &$ -1/3$ &$1/6 $ &$ 7/6$ &$-5/6$ &$2/3$ &$-1/3$ \\
      \hline
    \hline
   \end{tabular*}
\end{center}
 \end{table}
In these representations, the Yukawa and the relevant interaction terms  between the vector-like quarks and SM quarks are 
\begin{eqnarray}
{\cal L}_{SM}&=& -y_u {\bar q}_LH^c u_R -y_d{\bar q}_L H d_R \nonumber \\
{\cal L}_{{\cal U}_1, {\cal D}_1}&=& -y_T{\bar q}_LH^c U_{1_R} -y_B{\bar q}_L H D_{1_R}-m_T {\bar U}_L U_R-M_D {\bar D}_L D_R, \nonumber \\
{\cal L}_{{\cal D}_2}~~&=& -y_T {\bar D}_{2_L} H^c u_{R} -y_B{\bar D}_{2_L} H d_{R}-M_D {\bar D}_{2_L}  D_{2_R}, \nonumber \\
{\cal L}_{{\cal D}_X, {\cal D}_Y}&=& -y_T {\bar D}_{X_L}H u_{R} -y_B{\bar D}_{Y_L} H^c d_{R}-M_X {\bar D}_{X_L} D_{X_R}-M_Y {\bar D}_{Y_L} D_{Y_R}, \nonumber \\
{\cal L}_{{\cal T}_X, {\cal T}_Y}&=& -y_T {\bar q}_{L}\tau^a H^c  {\cal T}^a_{X_R} -y_B{\bar q}_{L} \tau^a H {\cal T}^a_{Y_R}-M_X {\bar  {\cal T}}_{X_L}  {\cal T}_{X_R}-M_Y {\bar  {\cal T}}_{Y_L}  {\cal T}_{Y_R}\, .
\end{eqnarray}
After spontaneous symmetry breaking, the Yukawa interactions generate mixing between the SM quarks and the vector-like quarks at tree level. The singlet vector-like quark and the triplet vector-like quark exhibit  similar mixing patterns, while the doublet vector-like quarks have a different mixing pattern. To avoid conflicts with low energy  experimental data (flavor changing neutral interactions), we limit  the vector-like quarks mixing with the third generation of SM quarks only. This is reasonable also because of the large mass gap between vector-like fermions and the first two generations of quarks. The mixing patterns will be described below. 

The gauge eigenstate fields resulting from the mixing can be written in general as,
\begin{eqnarray}
{\cal T}_{L,R}=&\left(\begin{matrix}
t\\T\end{matrix}\right)_{L,R}\ \qquad 
{\cal B}_{L,R}=&\left(\begin{matrix} b\\B\end{matrix}\right)_{L,R}\ 
\,
\label{eq:gauge_eigenst_fields}
\end{eqnarray}
The mass eigenstate fields are denoted as $(t,_1, t_2)$ and $(b_1, b_2)$ and they are found through bi-unitary transformations,
\begin{eqnarray}
{\mathbf T}_{L,R}&=&\left(\begin{matrix} t_1 \\t_2\end{matrix}\right)_{L,R}=V_{L,R}^t \left(\begin{matrix} t\\T\end{matrix}\right)_{L,R}\nonumber \\
{\mathbf B}_{L,R}&=&\left(\begin{matrix} b_1\\b_2\end{matrix}\right)_{L,R}=V_{L,R}^b \left(\begin{matrix} b\\B\end{matrix}\right)_{L,R}
\, ,
\end{eqnarray}
where
\begin{equation}
V_{L,R}^{t}=\left(
\begin{matrix}
\cos\theta& -\sin\theta\\
\sin\theta &\cos\theta\end{matrix} 
\right)_{L,R}\, , \qquad
V_{L,R}^{b}=\left(
\begin{matrix}
\cos\beta & -\sin\beta\\
\sin\beta &\cos\beta \end{matrix}
\right)_{L,R}\, ,
\label{eq:rotation_matrices}
\end{equation}
In the following we abbreviate $\cos\theta_L^t\equiv c_L^t$,.... 
Through these rotations we obtain the diagonal mass matrices 
\begin{equation}
M^t_{diag}=V_L^t M^t (V_R^t)^\dagger=\left(\begin{matrix}m_{t_1} & 0 \\0 &m_{t_2}\end{matrix}\right)
\quad , \quad
M^b_{diag}=V_L^b M^b (V_R^b)^\dagger=\left(\begin{matrix}m_{b_1} & 0 \\0 &m_{b_2}\end{matrix}\right)\, .
\end{equation}
the eigenvectors now become, for instance for the top sector
 \begin{eqnarray}
 m_{t_1,t_2}^2=\frac14 \left [ ( y_t^2+y_T^2)v^2+y_M^2u^2 \right] \left[ 1\pm \sqrt{1-\left( \frac{2y_t y_M vu}{ (y_t^2+y_T^2)v^2+y_M^2u^2}\right )^2} \right]
 \label{eq:eigenvecTt}
 \end{eqnarray}
 with eigenvectors
 \begin{equation}
\left(  \begin{array}{c} t_1 \\ t_2  \end{array} \right )_{L,R}=  \left(  \begin{array}{cc} \cos \theta_{L,R} & \sin \theta_{L,R} \\ -\sin \theta_{L,R}  & \cos \theta_{L,R} \end{array}  \right)
\left( \begin{array}{c} t \\T  \end{array} \right )_{L,R}
\, .
 \end{equation}

 where the mixing angles are
 \begin{eqnarray}
 \label{eq:mixing}
 \tan \theta_L&=&\frac{2y_T y_Mvu}{ y_M^2u^2-(y_t^2+y_T^2)v^2}\nonumber \\
\tan \theta_R&=& \frac{2y_t y_T v^2}{y_M^2u^2 -(y_t^2+y_T^2)v^2}\, ,
\end{eqnarray}
and similarly for the $b$-quark sector, with the replacement $t \to b$ and $\theta \to \beta$.
Note that, because of their charge assignments, the $X$ and $Y$
fields do not mix with the other fermions and are therefore also mass eigenstates.

Relationships between mixing angles and mass eigenstates 
depend on the representation \cite{Aguilar-Saavedra:2013qpa, Chen:2017hak}.
 \begin{eqnarray}
{\rm For~ doublets:} 
&(XT): &m_X^2=m_T^2 (\cos \theta_R)^{2}+m_t^2 (\sin \theta_R)^{2}\nonumber \\
&(TB): &m_T^2 (\cos \theta_R)^{2}+m_t^2 (\sin \theta_R)^{2}=m_B^2 (\cos \beta_R)^{2}+m_b^2 (\sin \beta_R)^{2}
\nonumber \\
&(BY):&m_Y^2=m_B^2 (\cos \beta_R)^{2}+m_b^2 (\sin \beta_R)^{2}\nonumber \\
\nonumber \\
{\rm For ~triplets:}& (XTB): &m_X^2=m_T^2 (\cos \theta_L)^{2}+m_t^2 (\sin \theta_L)^{2}\nonumber \\
 && \phantom{m_X^2} =m_B^2 (\cos \beta_L)^{2} + m_b^2(\sin \beta_L)^{2} \, ,\nonumber \\
  &&{\rm where}~\sin(2\beta_L)= \sqrt{2}{m_T^2-m_t^2\over  (m_B^2-m_b^2)}\sin(2\theta_L)\, .\nonumber \\
  &(TBY):& m_Y^2=m_B ^2 (\cos \beta_L)^2+m_b^2 (\sin \beta_L)^{2}\nonumber \\
  &&\phantom{m_Y^2}=m_T^2 (\cos \theta_L)^{2}+m_t^2 (\sin \theta_L)^{2} \, , \nonumber\\
 &&{\rm where}~\sin(2\beta_L)= {m_T^2-m_t^2\over \sqrt{2} (m_B^2-m_b^2)}\sin(2\theta_L)\, , 
 \label{relations}
\end{eqnarray}
and where 
\begin{eqnarray}
m_{T} \tan \theta_R&=m_{t}\tan \theta_L\qquad &{\hbox{for~singlets,~triplets}}\nonumber \\
m_{T} \tan \theta_L&=m_{t} \tan \theta_R\qquad &{\hbox{for~doublets}} \nonumber\\
m_{B} \tan \beta_R&=m_{b}\tan \beta_L\qquad &{\hbox{for~singlets,~triplets}}\nonumber \\
m_{B} \tan \beta_L&=m_{b} \tan \beta_R\qquad &{\hbox{for~doublets}}\, .
\label{angles}
\end{eqnarray}

For doublet models, while the Higgs mixing is the same as in the previous section,  the mixing between the top quark $t$ and the new vector-like singlet $T$, characterized by the mixing $\theta_L$, results in the shift in the Yukawa couplings as follows
\begin{eqnarray}
y_t(\mu_0)&=&\frac{\sqrt{2} m_t}{v}\frac{1}{\sqrt{\cos^2 \theta_L+x_t^2 \sin^2 \theta_L}}\, , \nonumber \\
y_T(\mu_0)&=&\frac{\sqrt{2} m_T}{v}\frac{\sin \theta_L \cos \theta_L (1-x_t^2)}{\sqrt{\cos^2 \theta_L+x_t^2 \sin^2 \theta_L}}\, , \nonumber \\
y_B(\mu_0)&=&\frac{\sqrt{2} m_B}{v}\frac{\sin \theta_L \cos \theta_L (1-x_b^2)}{\sqrt{\cos^2 \theta_L+x_t^2 \sin^2 \theta_L}}\, , \nonumber \\
y_M(\mu_0)&=&\frac{m_T+m_B}{\sqrt{2} u}\sqrt{\cos^2 \theta_L+x_t^2 \sin^2 \theta_L}\, ,
\end{eqnarray}
with $x_b=m_b/m_B$, and as before $x_t=m_t/m_T$. We use these expressions as initial conditions to the RGE equations, Eq. \ref{eq:rge1singletTBscalar}. We review the mass bounds on vector-like quarks, then proceed with our numerical analysis in Sec. \ref{subsec:analysis}.

\subsection{Experimental bounds on vector-like quark masses}
\label{subsec:VLQbounds}

 Searches for vector-like quarks have been performed at the LHC and various limits exist \cite{ATLAS-CONF-2016-101}, all obtained assuming  specific decay mechanisms. The Run 2 results from the LHC have improved previous limits from Run 1 by about 500 GeV \cite{moriond2018}. All measurements assume top and down-type vector-like quarks  to decay into one of the final states involving $Z$, $W$ or Higgs bosons with 100\% branching ratios. So far, lower limits around 800 GeV have been obtained. The most recent search at ATLAS obtains, with 95\% C.L., lower limits on the $T$ mass of 870 GeV (890 GeV) for the singlet model, 1.05 TeV (1.06 TeV) for the doublet model, and 1.16 TeV (1.17 TeV) for the pure $Zt$ decay mode quark \cite{Aaboud:2017qpr}.    The current experimental lower bound on the mass of the down-type vector-like quark which mixes only with the third generation quark is around 730 GeV from Run 2 of the LHC  and  around 900 GeV from Run 1. The current lower bound for a vector-like quark which mixes with the light quarks is around 760 GeV from Run 1. In our analyses, we  set a lower limit on all masses of 800 GeV, to allow for the consideration of the largest parameter space.
 
 We proceed with analyzing the representations in turn, showing the effects of the additional fermions on the RGEs, and the mass and mixing angles constraints on the fermions and additional boson for each.   All the relevant RGE for the Yukawa couplings, couplings between the bosons, and gauge coupling constants are given in the Appendix \ref{sec:appendix}.
 
 \subsection{Numerical Analysis}
 \label{subsec:analysis}

\begin{figure}[htbp]
	\centering
	\begin{subfigure}{.33\textwidth}\hspace{-0.5cm}
		\includegraphics[height=1.6in]{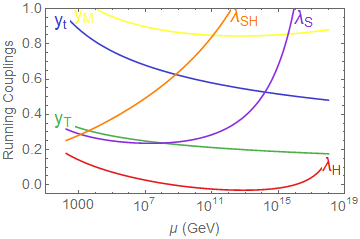}
	\end{subfigure}\hspace{-0.3cm}
	\begin{subfigure}{.33\textwidth}
		\includegraphics[height=1.6in]{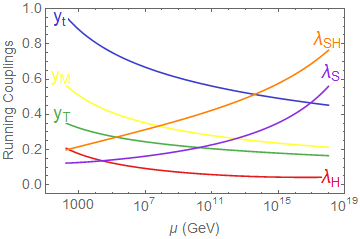}
	\end{subfigure}\hspace{0.2cm}
	\begin{subfigure}{.33\textwidth}
		\includegraphics[height=1.6in]{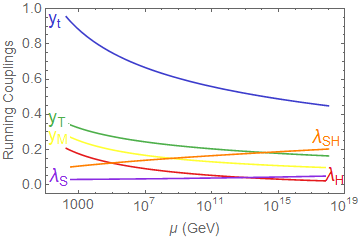}
	\end{subfigure}\\  
	\begin{subfigure}{.3\textwidth}\hspace{-1.2cm}
		\includegraphics[height=1.6in]{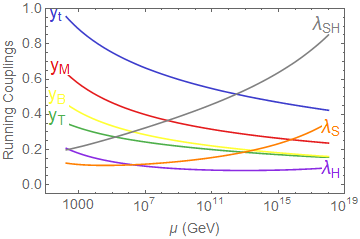}
	\end{subfigure}\hspace{-0.3cm}
	\begin{subfigure}{.33\textwidth}
		\includegraphics[height=1.6in]{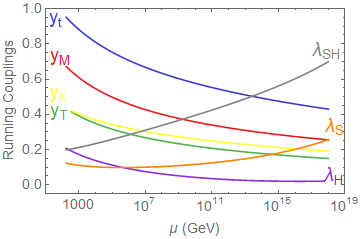}
	\end{subfigure}\hspace{0.1cm}
	  \caption{The RGE running of the Yukawa and scalar couplings for models with vector-like fermions. Top panel: singlet vector-like representations, ${\cal D}_1$ and ${\cal U}_1$ fermion models. We have set  $m_S=$ 0.8 TeV,  $u=$ 1 TeV, (left panel),  $m_S=$ 1 TeV and $u=2$ TeV (middle panel),  and $m_S=$ 1 TeV and $u=4$ TeV  (right panel).  Bottom panel: the same for the doublet vector-like models, for $u=2$ TeV,  (left panel) scalar+ vector-like $(T,B)$, (right panel) scalar+ vector-like $(X,T)$.   For the doublet representations we took: $m_T=0.8$ TeV, $m_B=1$ TeV, $m_X=1$ TeV,  $m_S=1$ TeV, $\mu_0=m_t$, and mixing angles  $\sin \varphi=0.1$ and $\sin \theta_L=0.08$.}
  \label{fig:rgeT}
\end{figure}

The evolution of the RGE's under different vector-like fermion representations  are illustrated in Fig. \ref{fig:rgeT} for different values of the VEVs of the new scalar field  ($u=1, 2, 4$ TeV). In the case of $u = 1$ TeV, we have taken the mass of the scalar
boson to be 0.8 TeV, because for $m_S=1$ TeV, the Higgs sector couplings diverge, leading to singularities, whereas
in the case of $u = 2$ TeV and $u = 4$ TeV,  we chose $m_S = 1$ TeV because
the smaller mass of 0.8 TeV is not large enough to ensure a positive Higgs quartic coupling.
 As required, all of the Higgs sector couplings remain positive up
to Planck scale. As expected, the fermion Yukawa couplings tend to decrease with increasing energy, while the
scalar bosonic couplings tend to increase. As we discussed previously, the addition of extra scalar
bosons to the Standard Model helps maintain a positive Higgs self-coupling at larger energy scales,
while the addition of extra fermions only aids in lowering it further. 
A common trend with respect to the models is that  the Yukawa couplings are generally negatively affected  by added loops at higher energy
scales, while the Higgs sector couplings are generally affected positively (they tend to increase with
increasing energy). 
\newpage
The obvious exception here is the SM Higgs coupling, which strays dangerously
close to zero at high energy scales, and even becomes negative for the additional singlet vector-like fermion case. The  models that augment the scalar boson by vector-like representations vary significantly among each other in predictions for the various couplings with the scalar VEV $u$. Note in particular that for the the first case, for the singlet vector-like case, where $u=1$, the Higgs couplings increase, and $\lambda_H$ becomes negative at $\mu \sim 10^{11}$, rendering the theory unstable, while if the additional scalar VEV  is increased to $u=2,$ or $4$ TeV, the problem is ameliorated. The same problem recurs for the doublet and triplet models (not shown), but the theory is safe for $u=2$ and $4$ TeV. Differences in the running RGE's are more pronounced for $\lambda_S$, the new boson self-coupling, and negligible for the others. Note in particular, the difference between the values in Fig. \ref{fig:rgeT} and \ref{fig:rgeS}.   For the doublet and triplet vector-like fermions, the RGE evolutions are similar,  and the Higgs self-coupling remains positive at all energies. The additional scalar quartic coupling does not  lie close to the
origin as its interactions with fermions are small.  There are some variations among models in the new scalar coupling, and the one describing the mixing with the SM Higgs. We have put  less emphasis
 on the  vacuum stability bound for the additional scalar, since its mass and
VEV are unknown, and thus  limiting concrete information from a detailed study of
its vacuum stability bound.

 Imposing the same  conditions on the positivity of the potential as in Eq. \ref{eq:singletstability}, we study the allowed masses and mixing angles  corresponding to each vector-like fermion representation.  In Figs. \ref{fig:fermionscalar_masses} and \ref{fig:3fermionscalar_masses}  we perform a scan over random values of the relevant vector-like quarks between 300 and 2200 GeV, and for the mixing angles $\sin \beta_L$ (in the bottom sector) and $\sin \theta_L$ (in the top sector) between -1 and 1. The allowed values of the scalar mass $m_S$ are plotted against the mixing angle in the scalar sector, $\sin \varphi$  for different values of the expectation values $u$ (1, 2 and 4 TeV), providing  an illustration of the possible quantitative properties of the
scalar boson in this model. The results are given for all models. In Fig. \ref{fig:fermionscalar_masses}  we plot the scans for singlet vector-like $T$ (top row), singlet vector-like $B$ (second row), $(T,B)$ doublet (third row), $(X,T)$ doublet (fourth row), $(B, Y)$ doublet (bottom row). And in Fig. \ref{fig:3fermionscalar_masses}  we consider the $(X,T,B)$ triplet (top row), and $(T,B,Y)$ triplet (bottom row),  providing an illustration of the possible quantitative properties of the
scalar boson in these models.  We remark from Figs. \ref{fig:fermionscalar_masses} and \ref{fig:3fermionscalar_masses}  that just as in the SM extension containing only an extra
scalar boson, considered in the previous section, mass mixing between the two
scalar bosons is always required, and this mixing is significant, $\sin \varphi \ge 0.2$.  Also, consistent with previous discussions, increasing the VEV $u$ enlarges the parameter space, which is now quite restricted for $u=1$ TeV. As expected, the addition of vector-like fermions worsens the stability of the potential, but larger VEVs (mass scales) survive. The mixing in the singlet ${\cal U}_1$ model is the most effective counter-term addition, in fact pretty much ruling out the scenario where $u=1$ TeV (unless the additional scalar is light, 600-1000 GeV), while the other representations provide much milder bounds for the same VEV.

\begin{figure}[htbp]
	\centering
	\begin{subfigure}{.3\textwidth}\hspace{-1.5cm}
		\includegraphics[height=1.4in]{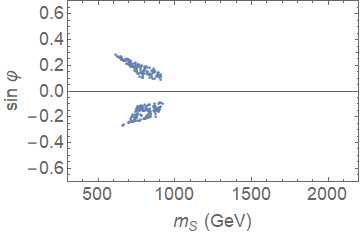}
		
	\end{subfigure}
	\begin{subfigure}{.3\textwidth}\hspace{-0.3cm}
		\includegraphics[height=1.4in]{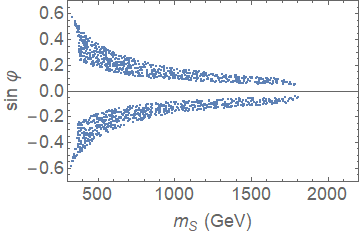}
		
	\end{subfigure}
	\begin{subfigure}{.3\textwidth}\hspace{0.4cm}
		\includegraphics[height=1.4in]{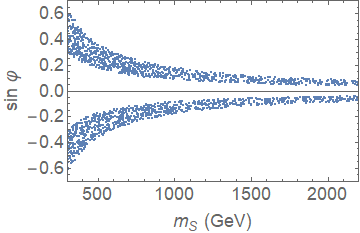}
		
	\end{subfigure}\\
	\begin{subfigure}{.3\textwidth}\hspace{-1.5cm}
		\includegraphics[height=1.4in]{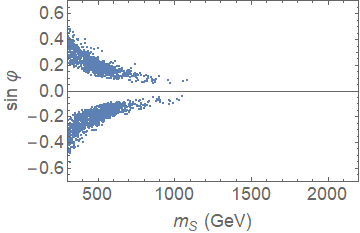}
		
	\end{subfigure}
	\begin{subfigure}{.3\textwidth}\hspace{-0.3cm}
		\includegraphics[height=1.4in]{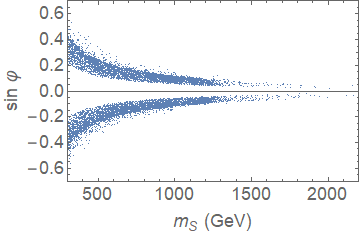}
		
	\end{subfigure}
	\begin{subfigure}{.3\textwidth}\hspace{0.4cm}
		\includegraphics[height=1.4in]{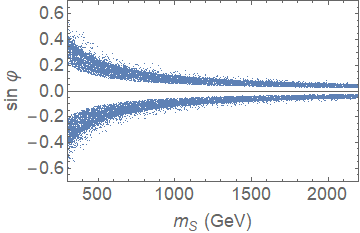}
		
	\end{subfigure}\\
	\begin{subfigure}{.3\textwidth}\hspace{-1.5cm}
		\includegraphics[height=1.4in]{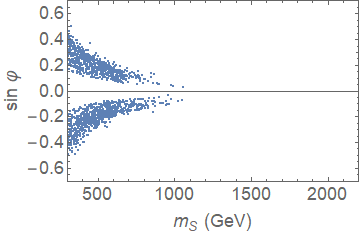}
		
	\end{subfigure}
	\begin{subfigure}{.3\textwidth}\hspace{-0.3cm}
		\includegraphics[height=1.4in]{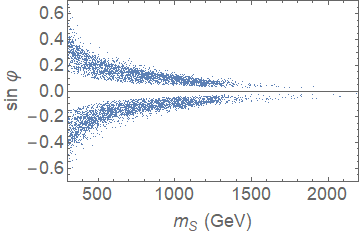}
		
	\end{subfigure}
	\begin{subfigure}{.3\textwidth}\hspace{0.4cm}
		\includegraphics[height=1.4in]{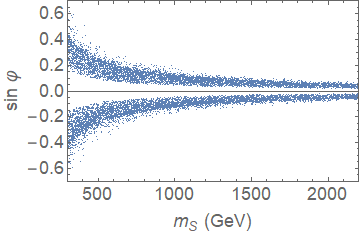}
		
	\end{subfigure}\\
	\begin{subfigure}{.3\textwidth}\hspace{-1.5cm}
		\includegraphics[height=1.4in]{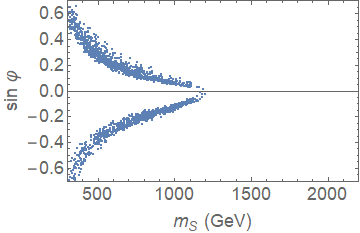}
		
	\end{subfigure}
	\begin{subfigure}{.3\textwidth}\hspace{-0.3cm}
		\includegraphics[height=1.4in]{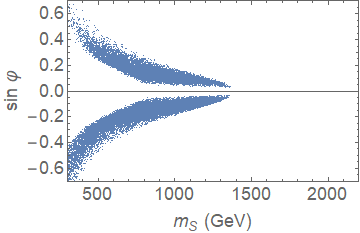}
		
	\end{subfigure}
	\begin{subfigure}{.3\textwidth}\hspace{0.4cm}
		\includegraphics[height=1.4in]{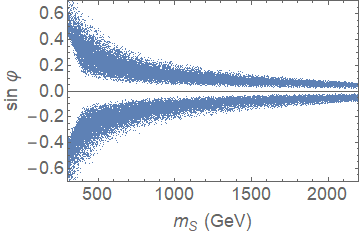}
		
	\end{subfigure}\\
	\begin{subfigure}{.3\textwidth}\hspace{-1.5cm}
		\includegraphics[height=1.4in]{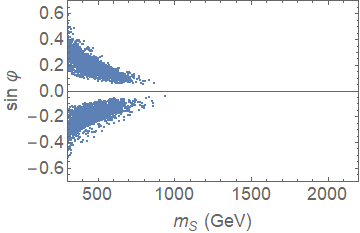}
		
	\end{subfigure}
	\begin{subfigure}{.3\textwidth}\hspace{-0.3cm}
		\includegraphics[height=1.4in]{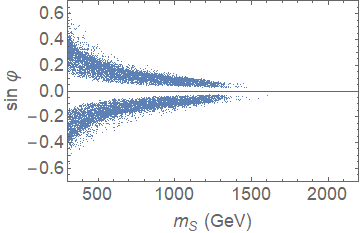}
		
	\end{subfigure}
	\begin{subfigure}{.3\textwidth}\hspace{0.4cm}
		\includegraphics[height=1.4in]{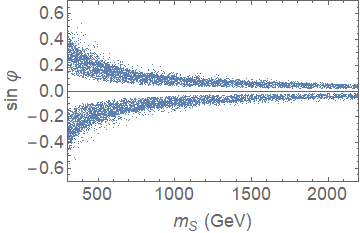}
		
	\end{subfigure}\\
	  \caption{The allowed parameter space for the scalar boson mass and mixing angle with the SM Higgs. In the  (top panel) scalar + vector-like $T$;   (second panel) scalar + vector-like $B$; (third panel) in the scalar + vector-like $(T,B)$ model; (fourth panel) scalar + vector-like $(X,T)$
fermion model; and (bottom panel) scalar + vector-like $(B,Y)$
fermion model, for different vacuum
expectation values of the additional scalar: $u=1$ TeV (left panel); $u=2$ TeV (middle panel); and $u=4$ TeV (right panel).}
  \label{fig:fermionscalar_masses}
\end{figure}
\begin{figure}[htbp]
	\centering
\begin{subfigure}{.3\textwidth}\hspace{-1.5cm}
		\includegraphics[height=1.4in]{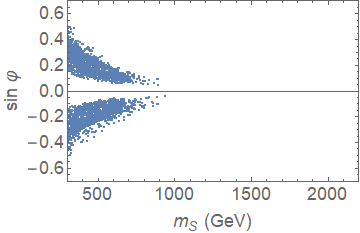}
		
	\end{subfigure}
	\begin{subfigure}{.3\textwidth}\hspace{-0.3cm}
		\includegraphics[height=1.4in]{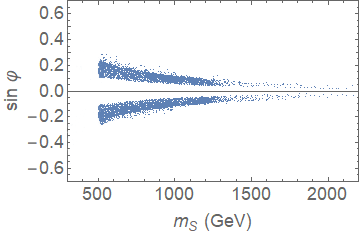}
		
	\end{subfigure}
	\begin{subfigure}{.3\textwidth}\hspace{0.4cm}
		\includegraphics[height=1.4in]{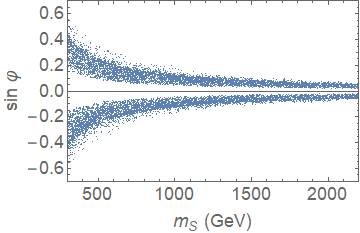}
		
	\end{subfigure}
	\begin{subfigure}{.3\textwidth}\hspace{-1.5cm}
		\includegraphics[height=1.4in]{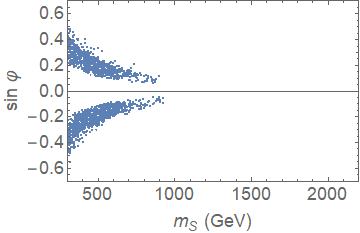}
		
	\end{subfigure}
	\begin{subfigure}{.3\textwidth}\hspace{-0.3cm}
		\includegraphics[height=1.4in]{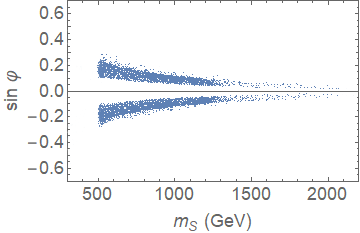}
		
	\end{subfigure}
	\begin{subfigure}{.3\textwidth}\hspace{0.4cm}
		\includegraphics[height=1.4in]{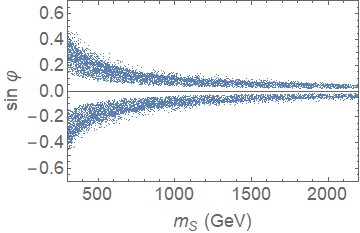}
		
	\end{subfigure}
	\caption{ (Continued) The allowed parameter space for the scalar boson mass and mixing angle with the SM Higgs. In the (top panel) the  scalar + triplet $(X, T, B)$ model , and for (bottom panel) the scalar + $(T,B,Y)$ triplet model, for different vacuum
expectation values of the additional scalar: $u=1$ TeV (left panel); $u=2$ TeV (middle panel); and $u=4$ TeV (right panel).}
\label{fig:3fermionscalar_masses}
\end{figure}
  
\begin{figure}[!h]
	\centering
	\begin{subfigure}{.3\textwidth}\hspace{-1.5cm}
		\includegraphics[height=1.4in]{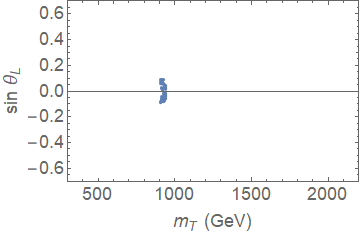}
		
	\end{subfigure}
	\begin{subfigure}{.3\textwidth}\hspace{-0.4cm}
		\includegraphics[height=1.4in]{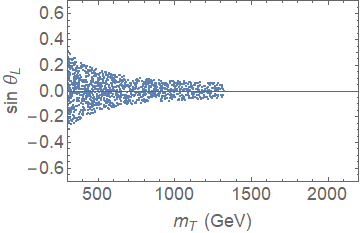}
		
	\end{subfigure}
	\begin{subfigure}{.3\textwidth}\hspace{0.5cm}
		\includegraphics[height=1.4in]{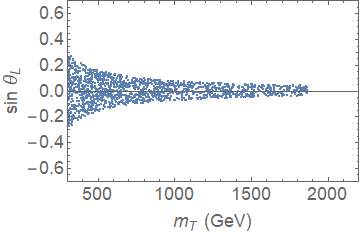}
		
	\end{subfigure}\\
	\begin{subfigure}{.3\textwidth}\hspace{-1.5cm}
		\includegraphics[height=1.4in]{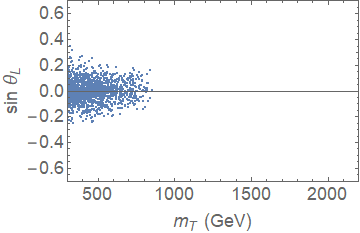}
		
	\end{subfigure}
	\begin{subfigure}{.3\textwidth}\hspace{-0.4cm}
		\includegraphics[height=1.4in]{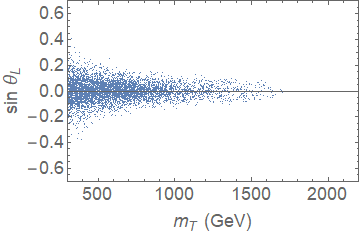}
		
	\end{subfigure}
	\begin{subfigure}{.3\textwidth}\hspace{0.5cm}
		\includegraphics[height=1.4in]{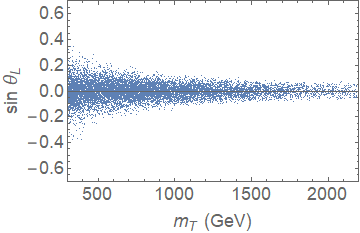}
		
	\end{subfigure}\\
	\begin{subfigure}{.3\textwidth}\hspace{-1.5cm}
		\includegraphics[height=1.4in]{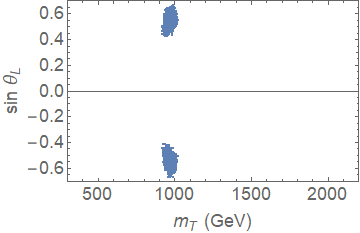}
		
	\end{subfigure}
	\begin{subfigure}{.3\textwidth}\hspace{-0.4cm}
		\includegraphics[height=1.5in]{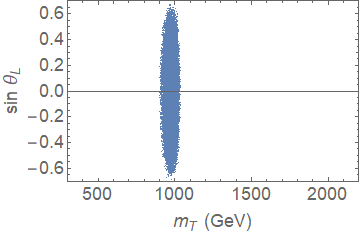}
		
	\end{subfigure}
	\begin{subfigure}{.3\textwidth}\hspace{0.5cm}
		\includegraphics[height=1.4in]{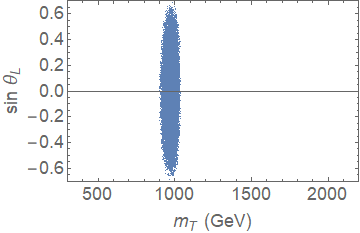}
		
	\end{subfigure}\\
	\begin{subfigure}{.3\textwidth}\hspace{-1.5cm}
		\includegraphics[height=1.5in]{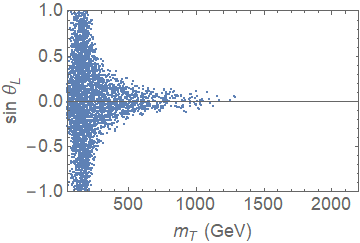}
		
	\end{subfigure}
	\begin{subfigure}{.3\textwidth}\hspace{-0.4cm}
		\includegraphics[height=1.5in]{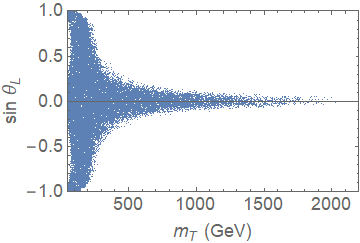}
		
	\end{subfigure}
	\begin{subfigure}{.3\textwidth}\hspace{0.5cm}
		\includegraphics[height=1.5in]{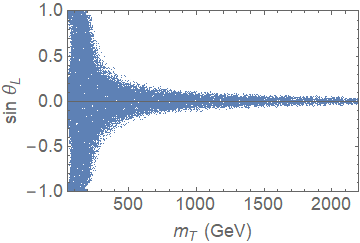}
		
	\end{subfigure}
	\begin{subfigure}{.3\textwidth}\hspace{-1.5cm}
		\includegraphics[height=1.5in]{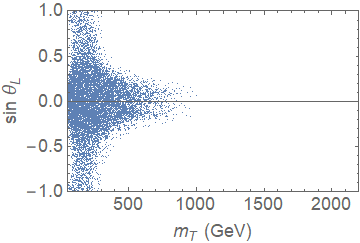}
		
	\end{subfigure}
	\begin{subfigure}{.3\textwidth}\hspace{-0.4cm}
		\includegraphics[height=1.5in]{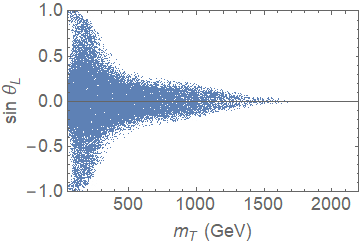}
		
	\end{subfigure}
	\begin{subfigure}{.3\textwidth}\hspace{0.5cm}
		\includegraphics[height=1.5in]{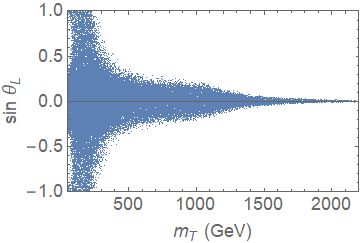}
		
	\end{subfigure}

	\caption{The allowed parameter space for the $T$ fermion mass and mixing angle with the top quark for: (top panel) in the scalar + singlet vector-like $T$ model; (second panel)  in scalar + vector-like $(T,B)$  model; (third panel)  for the $T$ fermion mass and mixing angle in the scalar + $(X,T)$ fermion doublet model, (fourth panel) for the scalar + $(X,T,B)$ triplet, and (bottom panel) for the triplet $(T,B,Y)$ model, for different vacuum expectation values, $u=1$ TeV (left panel); $u=2$ TeV (middle panel); and $u=4$ TeV (right panel).}
	\label{fig:fermionfermion_masses}
\end{figure}

 We now investigate the restrictions on the vector-like fermion masses and mixing from requiring the stability of the Higgs potential. 
We concentrate first on the vector-like $T$, which has the same change as the top quark, and which, through mixing can affect changes in the Higgs potential, both in the fermion and in the scalar sector. In
order to investigate this, we perform the same  scan over random values of $m_S$ and $M_T$ between 300 and 2200 GeV, and for the mixing angles $\sin \varphi$ and $\sin \theta_L$ between -1 and 1, and show the results in  Fig.   \ref{fig:fermionfermion_masses}. The first row shows the results for the singlet $T$ vector-like quark,  the second row shows the results for the doublet vector-like $(T, B)$, the third for the $(X,T)$ doublet, the fourth for the $(X,T,B)$ triplet and the last for the $(T,B,Y)$ triplet. 
Unlike the case of scalar mixing, here the mixing between the top quark and the vector-like one is required to be small, in general, for most cases, in the region $\sin \theta_L \in (-0.2, 0.2)$ (with some exceptions, where it can be larger, discussed below), and it can be zero.  The allowed masses of the $T$ quark are restricted for the scalar VEV $u=1$ TeV, and increase with  increasing  VEVs, so that in the singlet $T$ and doublet $(X,T)$ models cases, practically no $T$ masses are allowed for $u=1$ TeV, while masses up to 1400 GeV are allowed for $u=4$ TeV. For the SM like doublet $(T,B)$, for $u=1$ TeV, $m_T \le 800$ GeV, for $u=2$ TeV, $m_T \le 1600$ GeV, while for $u=4$ TeV, $m_T$ spans the whole axis. Note that here, like in the scalar sector, there are marked differences between the scenarios. For the doublet $(X, T)$, any mixing between the $T$ and $t$ quark is allowed. We expect this case to be somewhat similar to the singlet, however, the Yukawa coupling of the $X$ quark lowers the Higgs coupling further,  the parameter space continues to be severely constrained, and the mass is allowed in a narrow region near $m_T=1$ TeV for all values of the additional singlet. Here, as an exception to small mixing, the constraints on the mixing with the top are lifted. The case with triplets $(X,T,B)$, affected by both the $X$ and $B$ vector-like quarks, exhibits a behaviour independent of the singlet VEV. Masses again are favoured to be near $m_T=1$ TeV (we rule out light masses, $\sim 500$ from direct searches) and the mixing is allowed to be small or large. The strong enhancements are for the cases where the $t$ and $T$  mix. The mixing is expected to be stronger than between $B$ and $b$ quarks, due to the differences between mass of the top and of the bottom (making the denominator in Eq. \ref{eq:mixing} smaller). It is interesting to note here the effect of the $X$ vector-like quark, which, while not mixing with SM quarks, is nevertheless important for the mass of the $T$ vector-like quark (seen clearly if we compare the singlet $T$ model with the doublet $(X,T)$, and the doublet $(T,B)$ model with the triplet $(X,T,B)$). 

\begin{figure}[!h]
	\centering
	\begin{subfigure}{.3\textwidth}\hspace{-1.8cm}
		\includegraphics[height=1.5in]{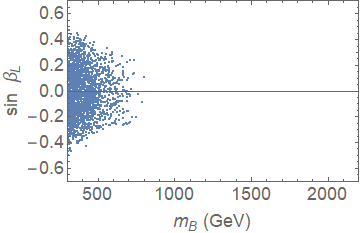}
		
	\end{subfigure}
	\begin{subfigure}{.3\textwidth}\hspace{-0.5cm}
		\includegraphics[height=1.5in]{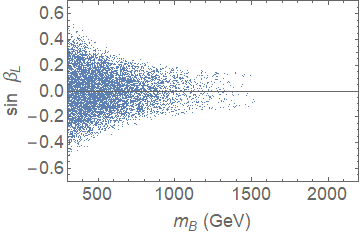}
		
	\end{subfigure}\hspace{0.6cm}
	\begin{subfigure}{.3\textwidth}
		\includegraphics[height=1.5in]{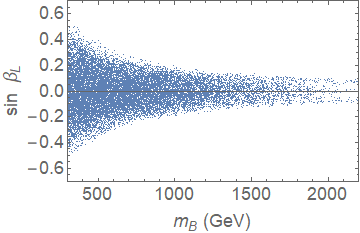}
		
	\end{subfigure}\\
	\begin{subfigure}{.3\textwidth}\hspace{-1.8cm}
		\includegraphics[height=1.5in]{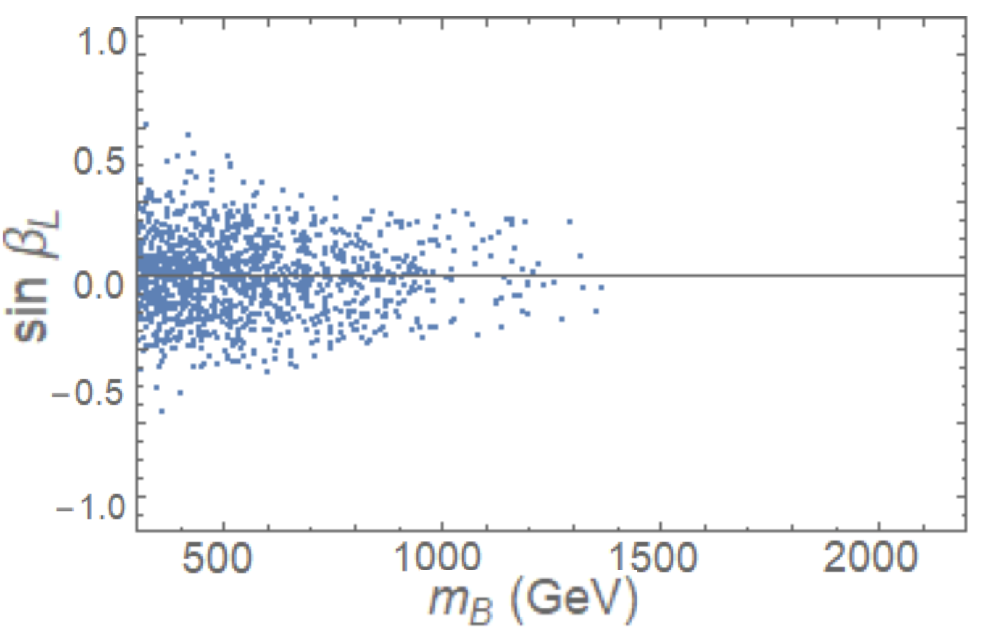}
		
	\end{subfigure}
	\begin{subfigure}{.3\textwidth}\hspace{-0.5cm}
		\includegraphics[height=1.5in]{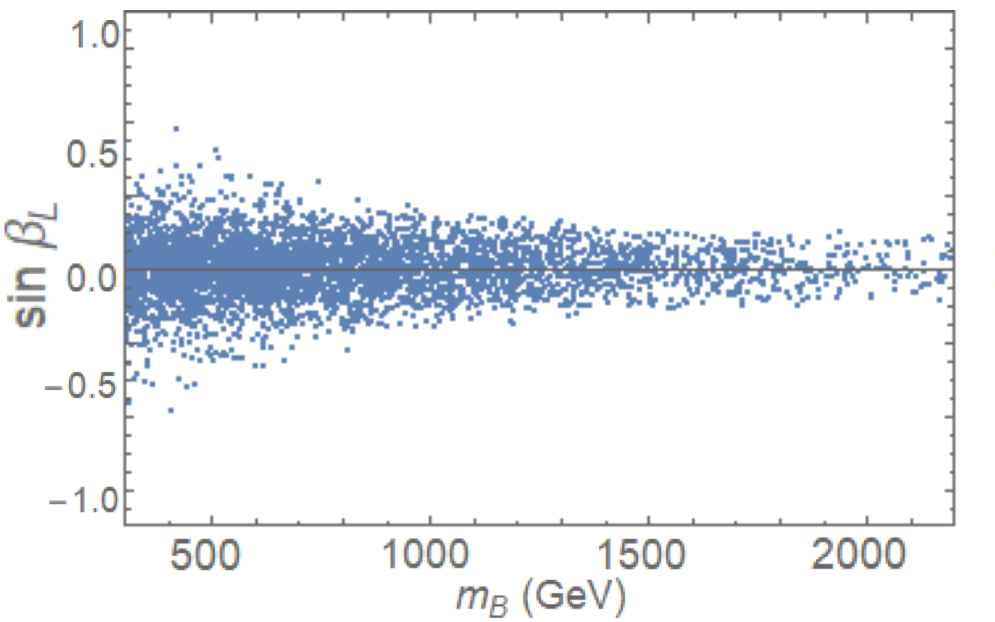}
		
	\end{subfigure}\hspace{0.6cm}
	\begin{subfigure}{.3\textwidth}
		\includegraphics[height=1.5in]{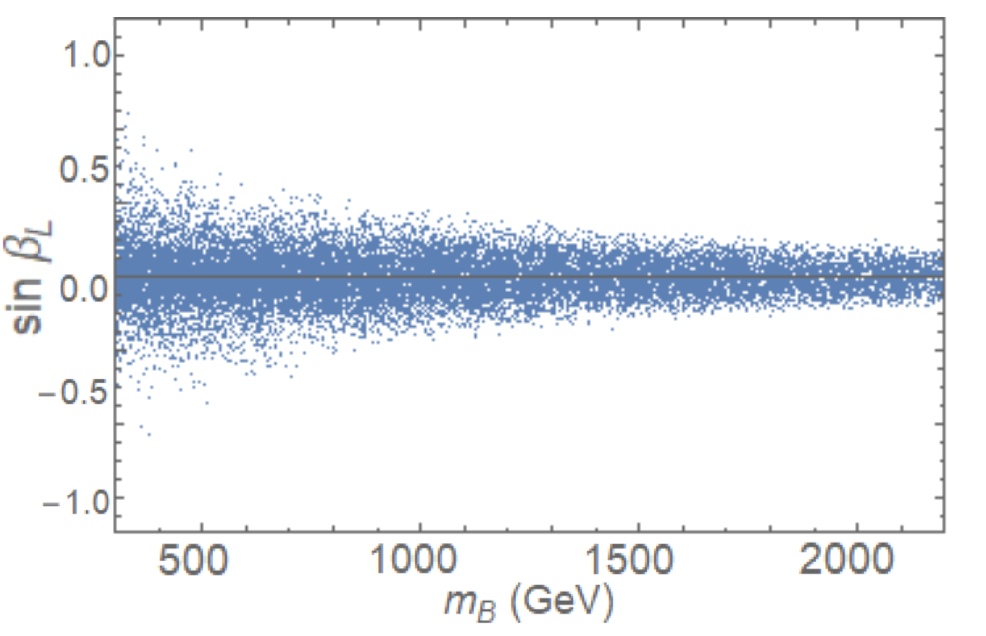}
		
	\end{subfigure}\\
	\begin{subfigure}{.3\textwidth}\hspace{-1.8cm}
		\includegraphics[height=1.5in]{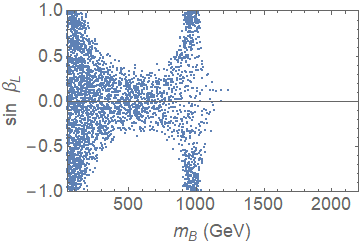}
		
	\end{subfigure}
	\begin{subfigure}{.3\textwidth}\hspace{-0.5cm}
		\includegraphics[height=1.5in]{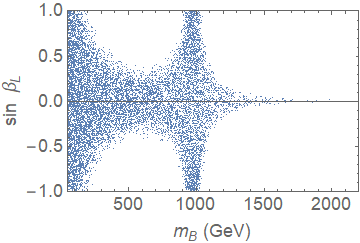}
		
	\end{subfigure}\hspace{0.6cm}
	\begin{subfigure}{.3\textwidth}
		\includegraphics[height=1.5in]{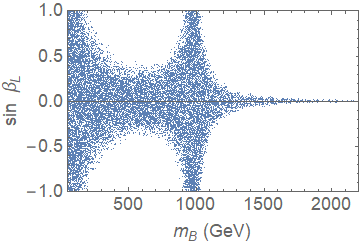}
		
	\end{subfigure}\\
	\begin{subfigure}{.3\textwidth}\hspace{-1.8cm}
		\includegraphics[height=1.5in]{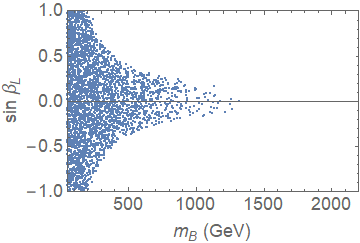}
		
	\end{subfigure}
	\begin{subfigure}{.3\textwidth}\hspace{-0.5cm}
		\includegraphics[height=1.5in]{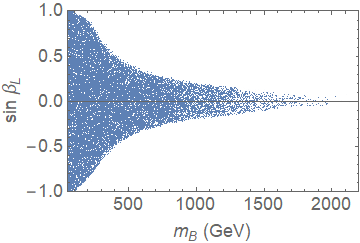}
		
	\end{subfigure}\hspace{0.6cm}
	\begin{subfigure}{.3\textwidth}
		\includegraphics[height=1.5in]{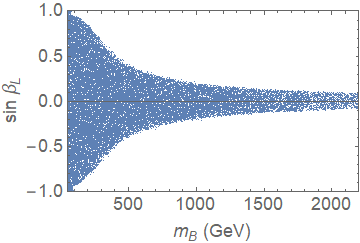}
		
	\end{subfigure}
	\begin{subfigure}{.3\textwidth}\hspace{-1.8cm}
		\includegraphics[height=1.5in]{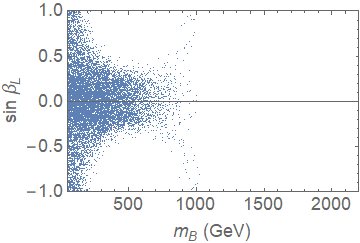}
		
	\end{subfigure}
	\begin{subfigure}{.3\textwidth}\hspace{-0.5cm}
		\includegraphics[height=1.5in]{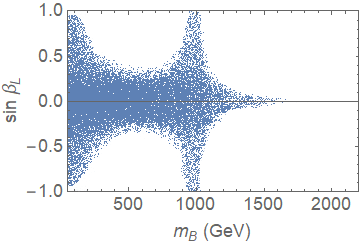}
		
	\end{subfigure}\hspace{0.6cm}
	\begin{subfigure}{.3\textwidth}
		\includegraphics[height=1.5in]{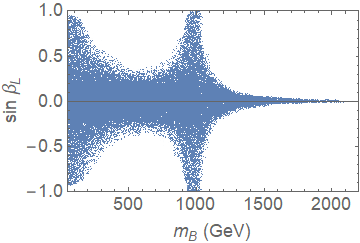}
		
	\end{subfigure}

  \caption{The allowed parameter space for the $B$ fermion mass and mixing angle in: (top panel) the  vector-like singlet $B$ model, (second panel) the  vector-like $(T,B)$  model, (third  panel) the  vector-like $(B,Y)$  model,  (fourth panel)  the  vector-like $(X,T,B)$ triplet, and (bottom panel) the  $(T,B,Y)$ triplet,  for different vacuum expectation values of the additional scalar: $u=1$ TeV (left panel); $u=2$ TeV (middle panel); and $u=4$ TeV (right panel).}
  \label{fig:fermionBfermion_masses}
\end{figure}

The scans in Fig. \ref{fig:fermionBfermion_masses} illustrate the  allowed masses and mixing angles of the $B$ vector-like quark with the bottom quark for the SM augmented by the additional scalar. We show, in the top panel,  the  vector-like singlet $B$ model, in the second panel, the vector-like $(T,B)$  model, in the third  panel, the  vector-like $(B,Y)$  model, in the fourth panel,  in the  vector-like $(X,T,B)$ triplet, and  in the bottom panel,  the   $(T,B,Y)$ triplet.  
  We again perform the same scan over the $m_B$ and $m_S$ masses and mixing angles $\sin \beta_L$ constrained by vacuum stability requirement, and plot the resulting $m_B$ against the mixing the $b$-sector $\sin \beta_L$.  The effect of the $B$ quark is markedly different from that of the $T$ quark due to weaker constraints on its angle (the denominator in $\tan \beta_L$ is larger than $\tan \theta_L$ ). For the  $B$ singlet model, the mixing and mass range are restricted, especially for $u=1$ TeV, while for the $(T,B)$ model the mass restrictions are lifted, but the mixing limits still remain. For the $(B,Y)$ doublet and for the $(T,B,Y)$ triplet model, the situation is very similar to that of the $T$ mass and mixing in these models: the mixing is restricted everywhere except around 1000 GeV, and this result is independent on the scalar VEV.

\begin{figure}[!h]
	\centering
	\begin{subfigure}{.3\textwidth}\hspace{-1.4cm}
		\includegraphics[height=1.5in]{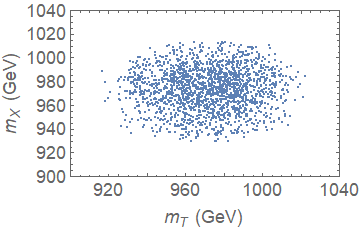}
		
	\end{subfigure}
	\begin{subfigure}{.33\textwidth}\hspace{-0.5cm}
		\includegraphics[height=1.5in]{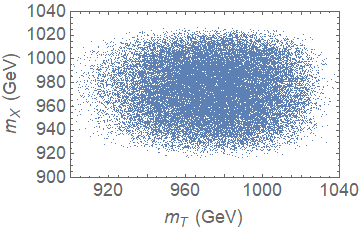}
		
	\end{subfigure}
	\begin{subfigure}{.3\textwidth}\hspace{-0.5cm}
		\includegraphics[height=1.5in]{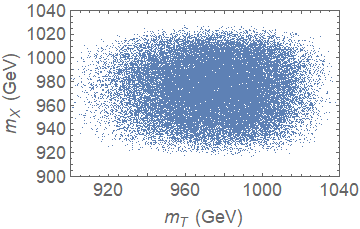}
		
	\end{subfigure}\\
	\begin{subfigure}{.3\textwidth}\hspace{-1.4cm}
		\includegraphics[height=1.4in]{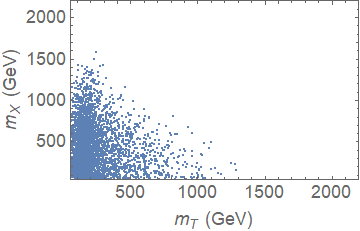}
		
	\end{subfigure}\hspace{0.1cm}
	\begin{subfigure}{.33\textwidth}\hspace{-0.5cm}
		\includegraphics[height=1.4in]{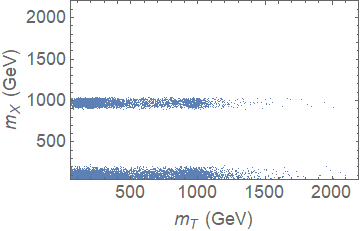}
		
	\end{subfigure}\hspace{0.1cm}
	\begin{subfigure}{.3\textwidth}
		\includegraphics[height=1.4in]{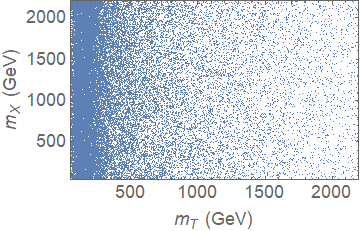}
		
	\end{subfigure}\\
	\begin{subfigure}{.3\textwidth}\hspace{-1.4cm}
		\includegraphics[height=1.4in]{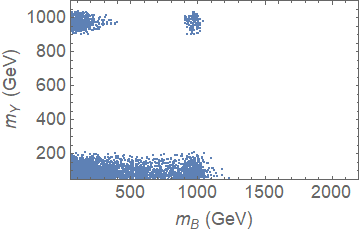}
		
	\end{subfigure}\hspace{0.1cm}
	\begin{subfigure}{.33\textwidth}\hspace{-0.5cm}
		\includegraphics[height=1.4in]{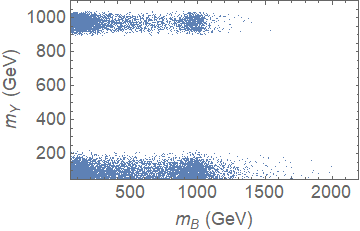}
		
	\end{subfigure}\hspace{0.1cm}
	\begin{subfigure}{.3\textwidth}
		\includegraphics[height=1.4in]{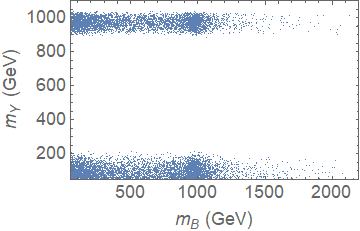}
		
		\end{subfigure}
		\begin{subfigure}{.3\textwidth}\hspace{-1.4cm}
		\includegraphics[height=1.4in]{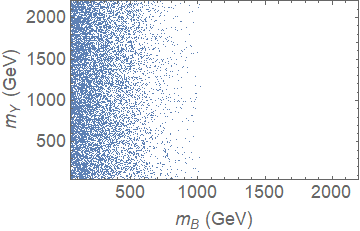}
		
	\end{subfigure}\hspace{0.1cm}
	\begin{subfigure}{.33\textwidth}\hspace{-0.5cm}
		\includegraphics[height=1.4in]{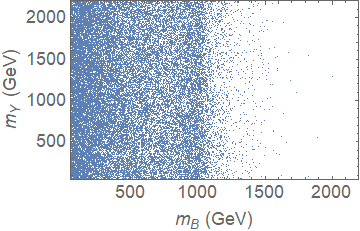}
		
	\end{subfigure}\hspace{0.1cm}
	\begin{subfigure}{.3\textwidth}
		\includegraphics[height=1.4in]{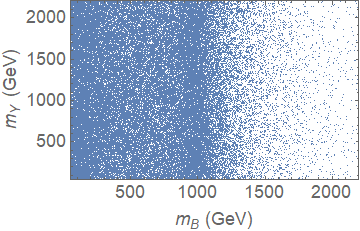}
		
	\end{subfigure}
\caption{The correlated parameter space for the $X$ and $T$ quark masses in the $(X,T)$ quark doublet model (top panel) and in the $(X,T,B)$ triplet model (second panel), and for the $Y$ and $B$ quark masses for the $(B,Y)$ doublet model (third panel), and for the $(T,B,Y)$ triplet model (bottom panel) for different vacuum expectation values.}
\label{fig:XTmasses}
\end{figure}

 Finally, we investigate constraints on the vector-like fermions with non-SM like hypercharge, $X$, with charge 5/3, and $Y$, with charge -4/3. As the additional vector-like quarks $X$ and $Y$ do not mix with SM particle, a plot of mass against the mixing angle does not make sense, Instead, in Fig.   \ref{fig:XTmasses}, the allowed values of the scanned fermion mass $m_X$ is plotted against  $m_T$, and $m_Y$ is correlated with $m_B$. Note that in the $(X, T)$ quark doublet, the $X$ and $T$ masses are strongly correlated (as seen from the third row of Fig. \ref{fig:fermionfermion_masses}) and the expected $m_X$ is required to be close to $1000$ GeV regardless of $m_T$ values. We see that, similarly, in the  $(B,Y)$ doublet model, $m_Y$ must have an allowed mass of  approximately 1000 GeV, regardless of  $m_B$, or the VEV $u$, unless both $m_X$ and $m_Y$ would be much lighter (100-200 GeV) in agreement with our earlier results. This seems to severely constrain models with vector-like quarks with exotic hypercharges.

For completeness, all the relevant RGE for the Yukawa couplings, couplings between the bosons and coupling constants are included in the Appendix \ref{sec:appendix}.

\newpage
\section{Electroweak Precision Measurements}
\label{sec:electroweakprecision}
Constraints on possible new physics also emerge from precision electroweak measurements. The extra singlet scalar and vector-like states induce modifications  to the vacuum polarizations of electroweak gauge bosons at loop level, which are parametrised by the oblique parameters $\mathbb{S, T},$ and $\mathbb{U}$.  For a large class of new physics models, corrections to precision electroweak observables are universal, in the sense that they are revealed only in self-energies of electroweak gauge bosons. There are solid constraints from these oblique corrections, pushing the scale of new physics around 1 TeV.  The oblique parameters can be calculated perturbatively for any model from the gauge boson propagators, and are defined as \cite{Hagiwara1995}.

\begin{eqnarray}
\label{eq:ewpgeneral}
\mathbb{S}&=&16\pi \Re\left[\bar{\Pi}^{3Q}_{\gamma}(m_{Z}^{2}) -\bar{\Pi}^{33}_{Z}(0)\right] ,\nonumber \\
\mathbb{T}&=&\frac{4\sqrt{2}G_{F}}{\alpha_{e}}\Re\left[\bar{\Pi}^{3Q}(0) -\bar{\Pi}^{11}(0)\right]   ,\nonumber \\
\mathbb{U}&=&16\pi \Re \left[\bar{\Pi}^{33}_{Z}(0) -\bar{\Pi}^{11}_{W}(0)\right] 
\end{eqnarray}
The current experimental values are obtained by fixing $\Delta \mathbb{U}=0$ are $\Delta \mathbb{T}=0.08\pm0.07$, $\Delta \mathbb{S}=0.05\pm0.09$. The overall calculation of $\mathbb{S, T}$ and $\mathbb{U}$ parameters via loop contributions can be separated into contributions from scalars and from fermions. Complete expressions for the contributions from the additional scalar and all fermion representations are given in the Appendix \ref{sec:appendix}.
  
\subsection{Contributions to the $\mathbb{S}$ and $\mathbb{T}$-parameters from scalar sector}
\label{subsec:scalarSTU}
\begin{figure}[!h]
	\centering
	\begin{subfigure}{.33\textwidth}\hspace{-1.5cm}
		\includegraphics[height=1.5in]{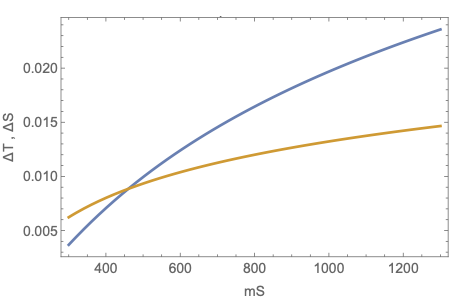}
		\caption{$\sin \varphi=$ 0.1}
	\end{subfigure}\hspace{-1.0cm}
	\begin{subfigure}{.33\textwidth}
		\includegraphics[height=1.5in]{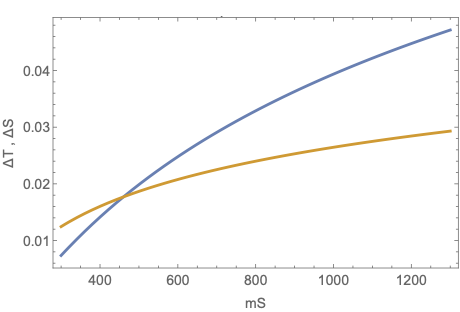}
		\caption{$\sin \varphi=$ 0.2}
	\end{subfigure}\hspace{0.3cm}
	\begin{subfigure}{.33\textwidth}
		\includegraphics[height=1.5in]{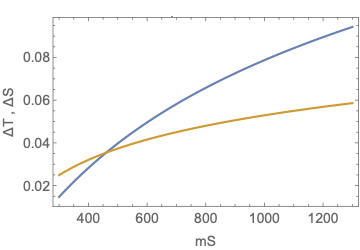}
		\caption{$\sin \varphi=$ 0.4}
	\end{subfigure}
  \caption{The contribution to the $\mathbb{T}$ (orange) and $\mathbb{S}$ (blue) parameters in the SM augmented by a singlet scalar, as a function of the singlet scalar mass. We take $u$ = 1 TeV for our consideration to remain in the vicinity of vacuum stability constraints.}
  \label{fig:scalarSTparam}
\end{figure}

Rewriting Eq. \ref{eq:ewpgeneral} explicitly in terms of the scalar loop contributions to the gauge boson two point functions
\begin{eqnarray}
\label{eq:ewpscalar}
\mathbb{S}_{SH}&=&\frac{16\pi}{m_{Z}^{2}}\Re\left[\frac{c_{w}^{2}}{eg_{z}}\left(\Pi_{z\gamma}(m_{Z}^{2})-\Pi_{z\gamma}(0) \right)+ \frac{s_{w}^{2}c_{w}^{2}}{e^{2}}\left(\Pi_{\gamma\gamma}(m_{Z}^{2})-\Pi_{\gamma\gamma}(0) \right) +\frac{1}{g_{z}^{2}}\left(\Pi_{ZZ}(m_{Z}^{2})-\Pi_{ZZ}(0) \right) \right] ,\nonumber \\
\mathbb{T}_{SH}&=&\frac{1}{\alpha_{e}}\Re\left[-\frac{\Pi_{WW}(0)}{m_{W}^{2}}+\frac{\Pi_{ZZ}(0)}{m_{Z}^{2}} + \frac{2s_{w}}{c_{w}}\frac{\Pi_{\gamma Z}(0)}{m_{Z}^{2}} +\frac{s_{w}^{2}}{c_{w}^{2}}\frac{\Pi_{\gamma\gamma}(0)}{m_{Z}^{2}} \right], \nonumber \\
\mathbb{U}_{SH}&=&16\pi \Re\left[\frac{1}{m_{Z}^{2}}\left(\frac{1}{g_{z}^{2}}\left(\Pi_{ZZ}(m_{Z}^{2})-\Pi_{ZZ}(0) \right)+ \frac{2s_{w}^{2}}{eg_{z}}\left(\Pi_{z\gamma}(m_{Z}^{2})-\Pi_{Z\gamma}(0) \right) +\frac{s_{w}^{4}}{e^{2}}\left(\Pi_{\gamma\gamma}(m_{Z}^{2})-\Pi_{\gamma\gamma}(0) \right) \right) \right. \nonumber \\
 &+& \left.\frac{1}{g^{2}m_{W}^{2}}\left(\Pi_{WW}(m_{W}^{2})-\Pi_{WW}(0) \right) \right]
\end{eqnarray}
Although pure scalar contributions to $\Delta \mathbb{S}$ and $\Delta \mathbb{T}$ relatively fit better with the experimental bounds as the scalar mixing angle is increased (Fig. \ref{fig:scalarSTparam}), we are particularly interested in numerical values at $\sin \varphi \sim 0.1$ and $m_{S}\sim$1 TeV since the constraints coming from vacuum stability are more restricted. Loop contributions from scalars to vector gauge boson two point functions are modified via scalar mixing and given in \cite{Kanemura:2015fra}.

Moreover, in Fig. \ref{fig:scalarSTparam}, it is seen that  the whole scalar parameter space $m_{S}, ~\sin \varphi$ is allowed, considering only the constraints from oblique parameters.

\subsection{VLQ contributions to the $\mathbb{S}$ and $\mathbb{T}$ parameters}
\label{subsec:fermionSTU}
The oblique correction parameter $\mathbb{T}$ for vector-like quarks is given as \cite{Lavoura:1992np}
\begin{eqnarray}
\mathbb{T}&=&\frac{N_{c}}{16\pi s_{w}^{2}c_{w}^{2}}\left[\sum_{\alpha,i}[(|V^{L}_{\alpha i}|^{2}+|V^{R}_{\alpha i}|^{2})\theta_{+}(y_{\alpha},y_{i})+2Re(V^{L}_{\alpha i}V^{R*}_{\alpha i})\theta_{-}(y_{\alpha},y_{i})] \right. \nonumber \\
 &-& \left.\sum_{\alpha,\beta}[(|U^{L}_{\alpha\beta}|^{2}+|U^{R}_{\alpha\beta}|^{2})\theta_{+}(y_{\alpha},y_{\beta})+2Re(U^{L}_{\alpha\beta}U^{R*}_{\alpha\beta})\theta_{-}(y_{\alpha},y_{\beta})]\right. \nonumber \\
 &-& \left. \sum_{i,j}[(|D^{L}_{ij}|^{2}+|D^{R}_{ij}|^{2})\theta_{+}(y_{i},y_{j})+2Re(D^{L}_{ij}D^{R*}_{ij})\theta_{-}(y_{i},y_{j})]   \right] \, ,
\end{eqnarray}
where the fermion ratio functions $\theta_{\pm}$ are given as
\begin{eqnarray}
\theta_{+}&=&y_{1}+y_{2}-\frac{2y_{1}y_{2}}{y_{1}-y_{2}}\ln\frac{y_{1}}{y_{2}}, \nonumber \\
\theta_{-}&=&2\sqrt{y_{1}y_{2}}\left(\frac{y_{1}+y_{2}}{y_{1}-y_{2}}\ln\frac{y_{1}}{y_{2}}-2 \right)
\end{eqnarray}
where $y_{i}=(\frac{m_{i}}{m_{Z}})^{2}$, and for the $\mathbb{S}$ parameter,
\begin{eqnarray}
\mathbb{S}&=&\frac{N_{c}}{2\pi}\left[\sum_{\alpha,i}[(|V^{L}_{\alpha i}|^{2}+|V^{R}_{\alpha i}|^{2})\psi_{+}(y_{\alpha},y_{i})+2Re(V^{L}_{\alpha i}V^{R*}_{\alpha i})\psi_{-}(y_{\alpha},y_{i})] \right. \nonumber \\
 &-& \left.\sum_{\alpha,\beta}[(|U^{L}_{\alpha\beta}|^{2}+|U^{R}_{\alpha\beta}|^{2})\chi_{+}(y_{\alpha},y_{\beta})+2Re(U^{L}_{\alpha\beta}U^{R*}_{\alpha\beta})\chi_{-}(y_{\alpha},y_{\beta})]\right. \nonumber \\
 &-& \left. \sum_{i,j}[(|D^{L}_{ij}|^{2}+|D^{R}_{ij}|^{2})\chi_{+}(y_{i},y_{j})+2Re(D^{L}_{ij}D^{R*}_{ij})\chi_{-}(y_{i},y_{j})]   \right]
\end{eqnarray}
Where $V_{\alpha i}^{L,R},U_{\alpha\beta}^{L,R}$ and $D_{ij}^{L,R}$ can be found in \cite{Aguilar-Saavedra:2013qpa}, and the functions $\psi_{\pm}$,$\chi_{\pm}$ are given respectively by
\begin{eqnarray}
\psi_{+}&=&\frac{22y_{\alpha}+14y_{i}}{9}-\frac{1}{9}\ln\frac{y_{\alpha}}{y_{i}}+\frac{1+11y_{\alpha}}{18}f(y_{\alpha},y_{\alpha})+\frac{7y_{i}-1}{18}f(y_{i},y_{i}), \nonumber \\
\psi_{-}&=&-\sqrt{y_{\alpha}y_{i}}\left(4+\frac{f(y_{\alpha},y_{\alpha})+f(y_{i},y_{i})}{2} \right) \, 
\end{eqnarray}
\begin{eqnarray}
\chi_{+}&=&\frac{y_{1}+y_{2}}{2}-\frac{(y_{1}-y_{2})^{2}}{3}+\left[ \frac{(y_{1}-y_{2})^{3}}{6}-\frac{1}{2}\frac{y_{1}^{2}+y_{2}^{2}}{y_{1}-y_{2}}\right]\ln\frac{y_{1}}{y_{2}}+\frac{y_{1}-1}{6}f(y_{1},y_{1})  \nonumber \\ 
&+&\frac{y_{2}-1}{6}f(y_{2},(y_{2})+\left[\frac{1}{3}-\frac{y_{1}+y_{2}}{6}-\frac{(y_{1}-y_{2})^{2}}{6} \right]f(y_{1},y_{2}) \nonumber \\
\chi_{-}&=&-\sqrt{y_{1}y_{2}}\left[2+(y_{1}-y_{2}-\frac{y_{1}+y_{2}}{y_{1}-y_{2}})\ln\frac{y_{1}}{y_{2}}+\frac{f(y_{1},y_{1})+f(y_{2},y_{2})}{2}-f(y_{1},y_{2}) \right] \, .
\end{eqnarray}
We now scan over different vector-like fermion masses via approximate expressions with only leading order terms \ref{sec:STapex}.
\begin{figure}[htbp]
	\centering
	\begin{subfigure}{.4\textwidth}\hspace{-1.5cm}
		\includegraphics[height=1.8in]{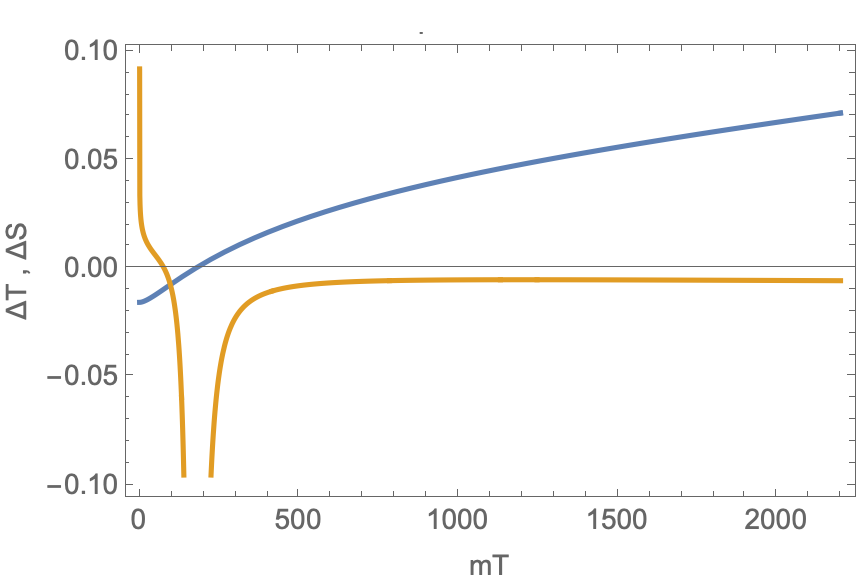}
		\caption{$\mathcal{U}_{1}$ (T)}
	\end{subfigure}\hspace{-0.2cm}
	\begin{subfigure}{.4\textwidth}
		\includegraphics[height=1.8in]{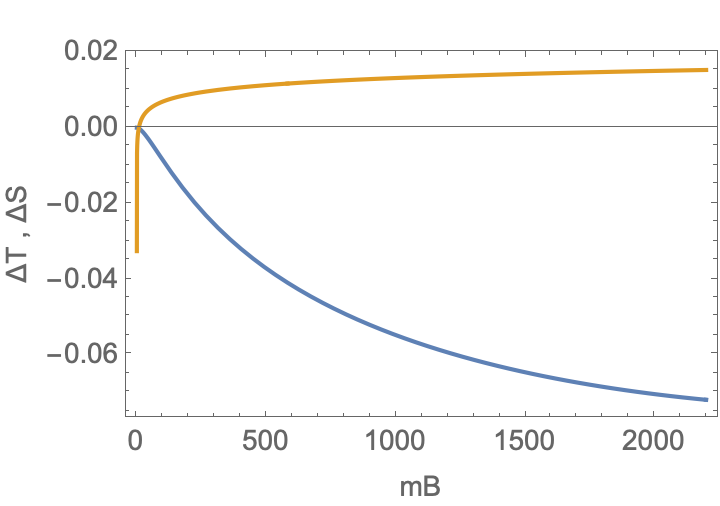}
		\caption{$\mathcal{D}_{1}$ (B)}
	\end{subfigure}
  \caption{The contributions to the $\mathbb{T}$ (blue) and $\mathbb{S}$ (orange) parameters in the singlet representations, as functions of the vector-like quark mass.}
  \label{fig:singletSTparam}
\end{figure}
\newpage
\begin{figure}[htbp]
	\centering
	\begin{subfigure}{.33\textwidth}\hspace{-1.5cm}
		\includegraphics[height=1.5in]{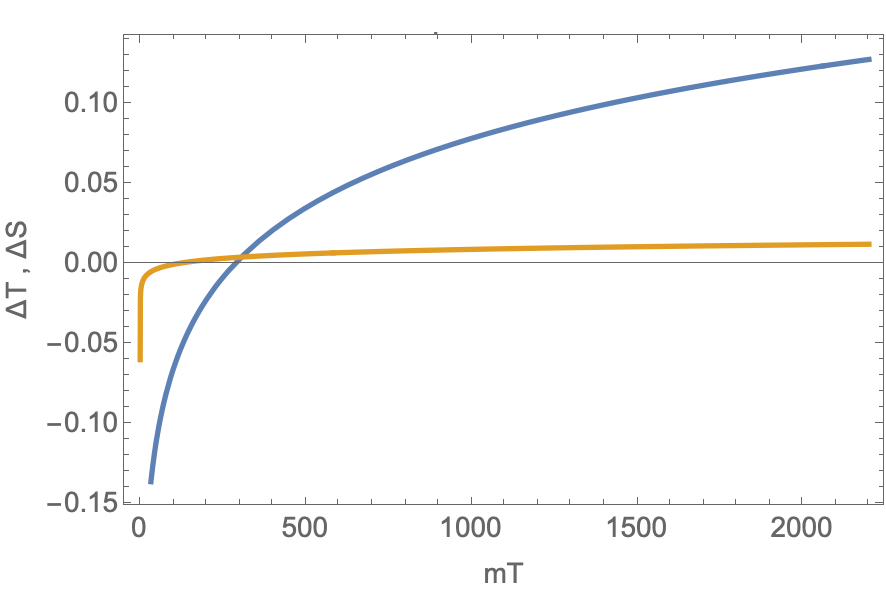}
		\caption{${\cal D}_2$ (T,B)}
	\end{subfigure}\hspace{-0.8cm}
	\begin{subfigure}{.33\textwidth}
		\includegraphics[height=1.5in]{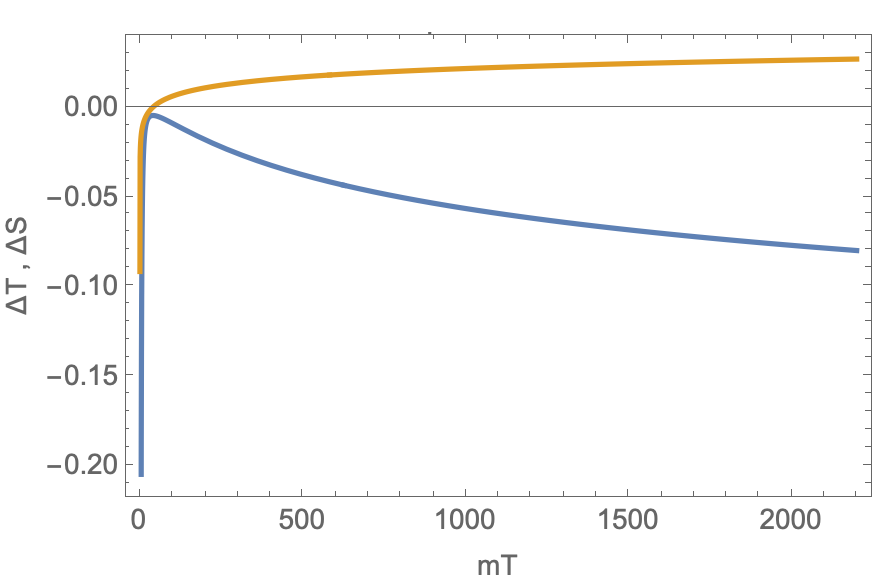}
		\caption{${\cal D}_X$ (X,T)}
	\end{subfigure}\hspace{0.3cm}
	\begin{subfigure}{.33\textwidth}
		\includegraphics[height=1.5in]{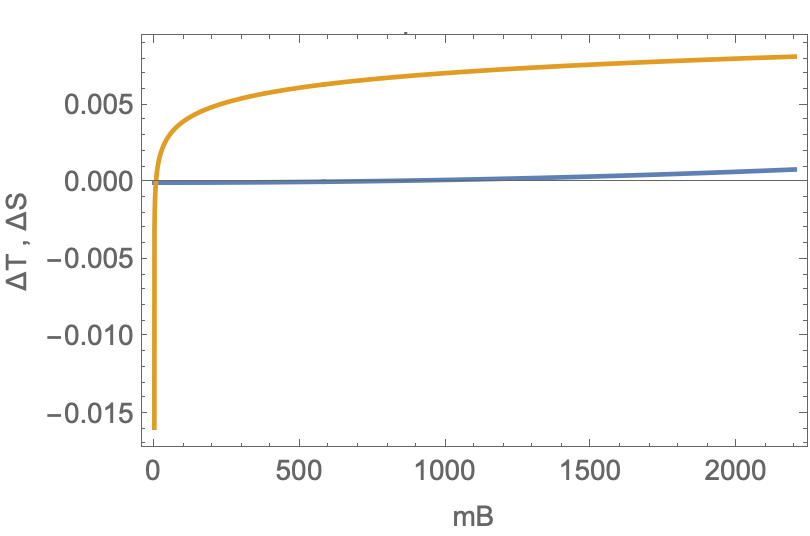}
		\caption{${\cal D}_Y$ (B,Y)}
	\end{subfigure}
  \caption{The contributions to the $\mathbb{T}$ (blue) and $\mathbb{S}$ (orange) parameters in the doublet representations as  functions of the vector-like quark mass.}
  \label{fig:doubletSTparam}
\end{figure}
\begin{figure}[htbp]
	\centering
	\begin{subfigure}{.4\textwidth}\hspace{-1.5cm}
		\includegraphics[height=1.8in]{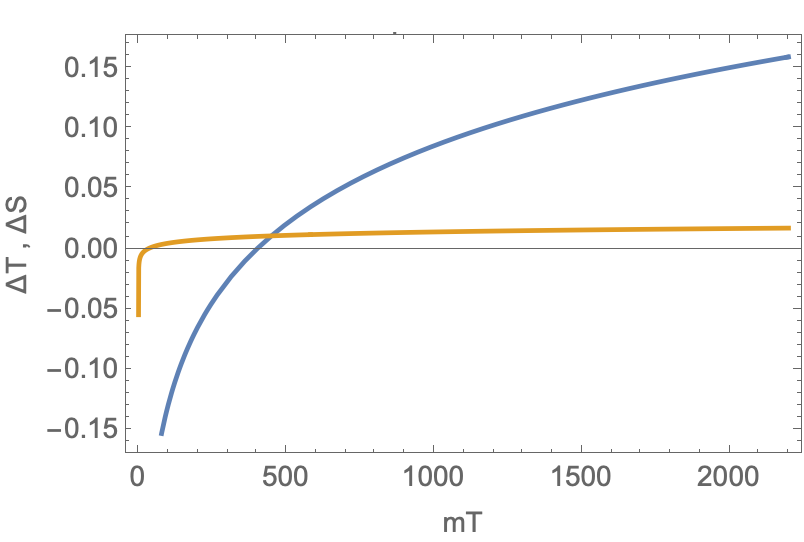}
		\caption{${\cal T}_X$ (X,T,B)}
	\end{subfigure}\hspace{-0.2 cm}
	\begin{subfigure}{.34\textwidth}
		\includegraphics[height=1.8in]{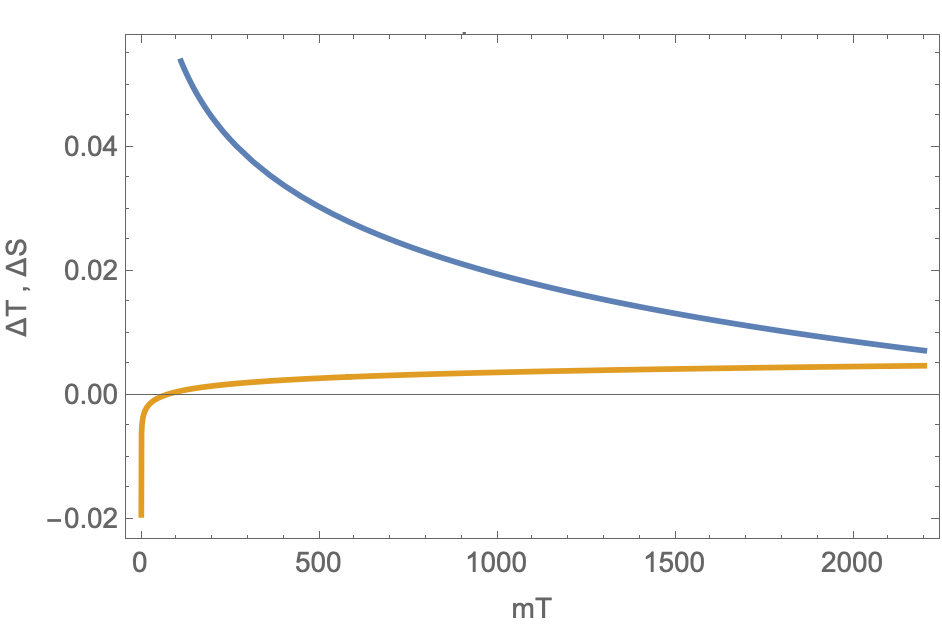}
		\caption{${\cal T}_Y$ (T,B,Y)}
	\end{subfigure}
  \caption{The contributions to the $\mathbb{T}$ (blue) and $\mathbb{S}$ (orange) parameters in the triplet representations, as  functions of the vector-like quark mass.}
  \label{fig:tripletSTparam}
\end{figure}

The $\mathbb{S}$-parameter agrees with the experimental bounds for small mixing angles, and does not bring  tighter constraints on the masses of vector-like quarks. However, the $\mathbb{T}$-parameter becomes negative for small mixing angles for the $\mathcal{D}_{1}$ and $\mathcal{D}_{X}$ representations. This feature in return  might exclude some regions of the parameter space once combined with the contributions from the SM + additional scalar, and imposes further conditions on the mass of singlet scalar. Apart from the vacuum stability constraints that connects the two sectors, this unique feature of electroweak precision accounts for the destructive interference between parameter spaces of scalars and vector-like fermions. Similar studies have been carried out in the literature \cite{Bahrami:2014ska} to impose more restricted constraints on parameter spaces of additional scalars. Checking the Eq.  \ref{eq:TSD1} for $\Delta \mathbb{T}$, the logarithmic term suppress the linear term in the small mixing domain of $\mathcal{D}_{1}$. Similarly, for $\mathcal{D}_{X}$, the first term in Eq. \ref{eq:TSDX} is inversely proportional to vector-like quark mass, which is rapidly suppressed by the second term, growing with  opposite sign with respect to mass of vector-like quark. Numerical values for  $\Delta \mathbb{T}$ and $\Delta \mathbb{S}$ at $m_{VLQ}\sim$ 1 TeV agree with the experimental limits in small mixing throughout all representations. However, it is shown from the mutual regions satisfying  $\mathbb{S, T}$-parameters and vacuum stability parameter spaces that, in general, vector-like quark masses and mixings are inversely proportional to each other. Moreover, in the mixing interval $\sin\theta_{L,R}>0.3$, although not shown here, $\Delta \mathbb{T}$ and $\Delta \mathbb{S}$ dangerously stray away from the experimental bounds, yielding more restrictions on $m_{S}$, as seen in Fig. \ref{fig:scalarSTparam}. Therefore, negative contributions to $\Delta \mathbb{T}$ and $\Delta \mathbb{S}$ are very likely to be compensated with relatively heavier scalars in various models.
\newpage

\section{Conclusions}
\label{sec:conclusion}

In this work, we presented a detailed analysis of the  stability conditions on the Higgs potential under the presence of extra vector-like fermions.  Since these have the same couplings for left and right components, they do not affect the loop-induced decays of the SM Higgs boson, and indeed, can have arbitrary bare masses in the Lagrangian.  We asked the question of whether they can have an effect on the Higgs sector, in particular, we concentrated on one of the outstanding problems in the SM, vacuum stability of the Higgs potential. While vector-like fermions appear in many beyond the SM models,  here we have taken a model-independent approach. We allowed mixing of the vector-like fermions with the third generation chiral fermions only, and we considered all possible anomaly free possibilities for the vector-like representations, with the additional fermions allowed to be in singlet, doublet, or triplet representations.

As all other fermions, their effect on the RGE's of the Higgs self-coupling constant is to lower it further, worsening the vacuum stability. An additional boson is introduced to alleviate this problem (representing an additional Higgs boson which would naturally appear in most New Physics models). Its presence is essential, and by itself it remedies the stability problem. The allowed additional scalar mass varies with its assigned VEV, but for all scenarios the mixing with the SM Higgs is required to be non-zero. 

Additional fermionic representations survive for scalar VEVs $u=1,2$ or $4$ TeV, and the agreement improves with increasing the scalar VEVs, indicating that higher scale physics is more likely to improve the vacuum stability problem. For most models, $u=1$ TeV is highly restricted, and likely ruled out, especially for top-like vector fermions, or in doublet models where this fermion is the only one mixing with the SM top. The situation worsens for the $(X,T)$ doublet model, where for all values of the VEV $u$, the mass $m_T$ hovers around 1000 GeV and is independent of the mixing. For triplet representations, the parameter spaces for $T-t$ mixing have similar characteristics in imposing a small vector-like quark mass limit, regardless of the value for the singlet VEV. Compared to the doublet $(T,B)$ representation, large mixing angles are permitted, for a relatively wide allowed mass spectrum. Comparatively,  the model $(X,T,B)$ is more sensitive to large vector-like quark masses, and shrinks the mixing angles to a small range as $m_T$ becomes large, whereas the model $(T,B,Y)$ allows for more parameter space for masses and mixing angles space for various singlet VEVs.  The differences in parameter spaces can be attributed to the fact that although the vector-like quark $X$ does not mix with the SM particles, its Yukawa term appears in the RGE for $y_T$, which is unique to the $(X,T,B)$ model.

Vacuum stability is improved if the bottom-like fermion is also present, and allowed to mix with the $b$ quark. The mixing angles are in general small (an exception are extreme cases where the mass is extremely restricted and the mixing completely free). However the difference between models with top-like or bottom-like quarks offer a way to distinguish between them, complimentary to collider searches.

Compared to $T$ vector-like quarks, constraints on the $B$-like fermion masses and mixing angles are much more relaxed. For the $(T,B)$ doublet model, the restrictions affect mostly $m_T$ and are relaxed for $m_B$ (the mixing with the $b$ quark is required to be small). While the mixing can be larger for the $(B,Y)$ and $(T,B,Y)$ models, and for $m_B=1000$ GeV, the mixing with the bottom quark is unrestricted. (On the other hand, the models (T,B,Y) and (B,Y) have almost identical parameter spaces for B mixings regardless of the singlet VEV. The mixings are constrained everywhere except $m_B=1000$ GeV. Although it’s possible, the model $(T,B,Y)$ is relatively less sufficient to impose mass values for $m_B$ around 1000 GeV for $u=1$ TeV. And finally, B mixings become less relaxed as the mass of vector-like quark gets larger for the models $(T,B)$ and $(X,T,B)$.) The  vector-like quarks carrying non-SM hypercharge do not mix with quarks, and seem to be required to have masses of around 1000 GeV, irrespective of the model, other vector-like fermion masses, or scalar VEV.

Compared to vacuum stability restrictions, electroweak precision constraints are more relaxed. Although the $\mathbb{S}$-parameter does not introduce strong restrictions on parameter space, the $\mathbb{T}$-parameter evolves in negative direction in different models. Combined with scalar contributions to $\mathbb{S}$ and $\mathbb{T}$-parameters, deviations from the experimental precision data might impose further restrictions on additional scalars and mixings with Higgs.

In conclusion, models where $T$-quark is unaccompanied by a $B$-quark yield very restrictive constraints for the masses $m_T$ and mixing angles $\sin \theta_L$. As well, additional vector-like fermions with hypercharge 5/3 or -4/3 are shown to restricted the additional fermions to masses close to 1 TeV, for the sampled range of the parameter space, which, in association with their exotic charges, renders them  predictable, making it easy to confirm or rule out the existence of these fermions. Our considerations which constrain the masses and mixings of vector-like fermions are complimented by analyses  on the parameters of models with vector-like quarks from  electroweak fits to the parameters  in these models..

\section{Acknowledgments}
We thank NSERC for partial financial support under grant number SAP105354.
 \newpage
 \section{Appendix}
 \label{sec:appendix}
 \subsection{RGEs for SM + additional boson + vector-like quarks}
We give below the renormalization group equations for the models studied in the text. The original expressions  appeared in \cite{RGEpt1}, and more expressions are included in \cite{Xiao:2014kba,Dhuria:2015ufo,Zhang:2015uuo}.
\subsubsection{Singlet ${\cal U}_1~(T)$, $Y=2/3$}
The relevant RGE for the Yukawa couplings are
\begin{eqnarray}
\label{eq:rge1singletTfermion}
\frac{dy_t^2}{d \ln \mu^2}&=& \frac{y_t^2}{16 \pi^2}\left (\frac{9y_t^2}{2}+\frac{9y_T^2}{2}-\frac{17g_1^2}{20}-\frac{9g_2^2}{4}- 8g_3^2 \right)\, ,\nonumber\\
\frac{dy_T^2}{d \ln \mu^2}&=& \frac{y_T^2}{16 \pi^2}\left (\frac{9y_t^2}{2}+ \frac{9y_T^2}{2}+\frac{y_M^2}{4}-\frac{17g_1^2}{20}-\frac{9g_2^2}{4}- 8g_3^2 \right)\, ,\nonumber\\
\frac{dy_M^2}{d \ln \mu^2}&=& \frac{y_M^2}{16 \pi^2}\left (y_T^2+\frac{9y_M^2}{2}-\frac{8g_1^2}{5}-8g_3^2\right).
\end{eqnarray}
The Higgs sector RGEs, describing the interactions between the two bosons are:
\begin{eqnarray}
\label{eq:rgeTsingletscalar}
\frac{d \lambda_H}{d \ln \mu^2}&=& \frac{1}{16 \pi^2} \left[  \lambda_H \left (12 \lambda_H+6 y_t^2+6 y_T^2-\frac{9 g_1^2}{10}-\frac{9 g_2^2}{2} \right)+ 
 \frac{\lambda_{SH}^2}{4}-3 y_t^4 -3 y_T^4-6y_t^2 y_T^2 \right. \nonumber \\
 &+& \left. \frac{ 27g_1^4}{100}+\frac{ 9g_2^4}{16} +\frac{ 9g_1^2 g_2^2 }{40}  \right ] \, , \nonumber \\
 \frac{d \lambda_S}{d \ln \mu^2}&=&\frac{1}{16 \pi^2} \left (9 \lambda_S^2+\lambda_{SH}^2 +12y_M^2\lambda_S-12y_M^4\right) \, , \nonumber \\
 \frac{d \lambda_{SH}}{d \ln \mu^2}&=&\frac{1}{16 \pi^2} \left [ \lambda_{SH} \left(2\lambda_{SH}+6 \lambda_H +3 \lambda_S +3 y_t^2  +3 y_T^2+6y_M^2-\frac{9g_1^2}{20}- \frac{9g_2^2}{4} \right)-12y_T^2y_M^2 \right ] ,
 \end{eqnarray}
\subsubsection{Singlet  ${\cal D}_1~(B)$, $Y=-1/3$}
The relevant RGE for the Yukawa couplings are
\begin{eqnarray}
\label{eq:rge1singletBfermion}
\frac{dy_t^2}{d \ln \mu^2}&=& \frac{y_t^2}{16 \pi^2}\left (\frac{9y_t^2}{2}+\frac{3y_B^2}{2}-\frac{17g_1^2}{20}-\frac{9g_2^2}{4}- 8g_3^2 \right)\, ,\nonumber\\
\frac{dy_B^2}{d \ln \mu^2}&=& \frac{y_B^2}{16 \pi^2}\left (\frac{3y_t^2}{2}+ \frac{9y_B^2}{2}+\frac{y_M^2}{4}-\frac{g_1^2}{4}-\frac{9g_2^2}{4}- 8g_3^2 \right)\, ,\nonumber\\
\frac{dy_M^2}{d \ln \mu^2}&=& \frac{y_M^2}{16 \pi^2}\left (y_B^2+\frac{9y_M^2}{2}-\frac{2g_1^2}{5}-8g_3^2\right).
\end{eqnarray}
\newpage
The Higgs sector RGEs, describing the interactions between the two bosons are:
\begin{eqnarray}
\label{eq:rge1singletscalar}
\frac{d \lambda_H}{d \ln \mu^2}&=& \frac{1}{16 \pi^2} \left[  \lambda_H \left (12 \lambda_H+6 y_t^2+6 y_B^2-\frac{9 g_1^2}{10}-\frac{9 g_2^2}{2} \right)+ 
 \frac{\lambda_{SH}^2}{4}-3 y_t^4 -3 y_B^4 \right. \nonumber \\
 &+& \left. \frac{ 27g_1^4}{100}+\frac{ 9g_2^4}{16} +\frac{ 9g_1^2 g_2^2 }{40}  \right ] \, , \nonumber \\
 \frac{d \lambda_S}{d \ln \mu^2}&=&\frac{1}{16 \pi^2} \left (9 \lambda_S^2+\lambda_{SH}^2 +12y_M^2\lambda_S-12y_M^4\right) \, , \nonumber \\
 \frac{d \lambda_{SH}}{d \ln \mu^2}&=&\frac{1}{16 \pi^2} \left [ \lambda_{SH} \left(2\lambda_{SH}+6 \lambda_H +3 \lambda_S +3 y_t^2  +3 y_B^2+6y_M^2-\frac{9g_1^2}{20}- \frac{9g_2^2}{4} \right)-12y_B^2y_M^2 \right ] ,
 \end{eqnarray}
Finally the coupling constants gain additional terms due to the new fermion, for both models ${\cal U}_1,~{\cal D}_1$ with singlet fermions as follows:
\begin{eqnarray}
\frac{dg_1^2}{d \ln \mu^2}&=&\frac{g_1^4}{16 \pi^2}\left(\frac{41}{10}+\frac{4}{15}\right )\, , \nonumber \\
\frac{dg_2^2}{d \ln \mu^2}&=&\frac{g_2^4}{16 \pi^2}\left( -\frac{19}{6} \right ) \, , \nonumber \\
\frac{dg_3^2}{d \ln \mu^2}&=&\frac{g_3^4}{16 \pi^2}\left(-7+\frac{2}{3}\right )\, .
\end{eqnarray}
\subsubsection{Doublet ${\cal D}_2$ $(T,B), \,Y=1/6$}
The relevant RGE for the Yukawa couplings are
\begin{eqnarray}
\label{eq:rge1singletTBfermion}
\frac{dy_t^2}{d \ln \mu^2}&=& \frac{y_t^2}{16 \pi^2}\left (\frac{9y_t^2}{2}+\frac{9y_T^2}{2}+\frac{3y_B^2}{2}+\frac{y_M^2}{2}-\frac{17g_1^2}{20}-\frac{9g_2^2}{4}- 8g_3^2 \right)\, ,\nonumber\\
\frac{dy_T^2}{d \ln \mu^2}&=& \frac{y_T^2}{16 \pi^2}\left (\frac{9y_t^2}{2}+ \frac{9y_T^2}{2}+\frac{3y_B^2}{2}+\frac{y_M^2}{2}-\frac{17g_1^2}{20}-\frac{9g_2^2}{4}- 8g_3^2 \right)\, ,\nonumber\\
\frac{dy_B^2}{d \ln \mu^2}&=& \frac{y_B^2}{16 \pi^2}\left (\frac{9y_t^2}{2}+ \frac{3y_T^2}{2}+\frac{9y_B^2}{2}+\frac{y_M^2}{2}-\frac{17g_1^2}{20}-\frac{9g_2^2}{4}- 8g_3^2 \right)\, ,\nonumber\\
\frac{dy_M^2}{d \ln \mu^2}&=& \frac{y_M^2}{16 \pi^2}\left (y_T^2+y_B^2+\frac{11y_M^2}{2}-\frac{g_1^2}{40}-\frac{9 g_2^2}{2}-8g_3^2\right).
\end{eqnarray}
The Higgs sector RGEs, describing the interactions between the two bosons are:
\begin{eqnarray}
\label{eq:rge1singletTBscalar}
\frac{d \lambda_H}{d \ln \mu^2}&=& \frac{1}{16 \pi^2} \left[  \lambda_H \left (12 \lambda_H+6 y_t^2+6 y_T^2+6y_B^2-\frac{9 g_1^2}{10}-\frac{9 g_2^2}{2} \right)+ 
 \frac{\lambda_{SH}^2}{4}-3 y_t^4 -3 y_T^4-3y_B^2- 6y_t^2 y_T^2 \right. \nonumber \\
 &-& 12 y_B^2 y_T^2  +\left.  \frac{ 27g_1^4}{400}+\frac{ 9g_2^4}{16} +\frac{ 9g_1^2 g_2^2 }{40}  \right ] \, , \nonumber \\
 \frac{d \lambda_S}{d \ln \mu^2}&=&\frac{1}{16 \pi^2} \left (9 \lambda_S^2+\lambda_{SH}^2 +12y_M^2\lambda_S-12y_M^4\right) \, , \nonumber \\
 \frac{d \lambda_{SH}}{d \ln \mu^2}&=&\frac{1}{16 \pi^2} \left [ \lambda_{SH} \left(2\lambda_{SH}+6 \lambda_H +3 \lambda_S +3 y_t^2  +3 y_T^2+3y_B^2+6y_M^2-\frac{9g_1^2}{20}- \frac{9g_2^2}{4} \right) \right. \nonumber \\
 &-&\left. 12y_T^2y_M^2 -12y_B^2y_M^2 \right ] \, , 
 \end{eqnarray}
\subsubsection{Doublet ${\cal D}_X$ $(X,T), \,Y=7/6$}
The relevant RGE for the Yukawa couplings are
\begin{eqnarray}
\label{eq:rge1singletTXfermion}
\frac{dy_t^2}{d \ln \mu^2}&=& \frac{y_t^2}{16 \pi^2}\left (\frac{9y_t^2}{2}+\frac{9y_T^2}{2}+\frac{9y_X^2}{2}+\frac{y_M^2}{2}-\frac{17g_1^2}{20}-\frac{9g_2^2}{4}- 8g_3^2 \right)\, ,\nonumber\\
\frac{dy_T^2}{d \ln \mu^2}&=& \frac{y_T^2}{16 \pi^2}\left (\frac{9y_t^2}{2}+\frac{9y_X^2}{2}+ \frac{9y_T^2}{2}+\frac{y_M^2}{2}-\frac{17g_1^2}{20}-\frac{9g_2^2}{4}- 8g_3^2 \right)\, ,\nonumber\\
\frac{dy_X^2}{d \ln \mu^2}&=& \frac{y_X^2}{16 \pi^2}\left (\frac{9y_t^2}{2}+ \frac{9y_X^2}{2}+ \frac{9y_T^2}{2}+\frac{y_M^2}{2}-\frac{17g_1^2}{20}-\frac{9g_2^2}{4}- 8g_3^2 \right)\, ,\nonumber\\
\frac{dy_M^2}{d \ln \mu^2}&=& \frac{y_M^2}{16 \pi^2}\left (y_T^2+y_X^2+\frac{11y_M^2}{2}-\frac{49g_1^2}{40}-\frac{9g_2^2}{2}-8g_3^2\right).
\end{eqnarray}
The Higgs sector RGEs, describing the interactions between the two bosons are:
\begin{eqnarray}
\label{eq:rgesingletTXscalar}
\frac{d \lambda_H}{d \ln \mu^2}&=& \frac{1}{16 \pi^2} \left[  \lambda_H \left (12 \lambda_H+6 y_t^2+6 y_T^2+6y_X^2-\frac{9 g_1^2}{10}-\frac{9 g_2^2}{2} \right)+ 
 \frac{\lambda_{SH}^2}{4}-3 y_t^4 -3 y_T^4-3y_X^2-6y_t^2 y_T^2 \right. \nonumber \\
 &-&\left.12 y_X^2 y_T^2+  \frac{ 27g_1^4}{400}+\frac{ 9g_2^4}{16} +\frac{ 9g_1^2 g_2^2 }{40}  \right ] \, , \nonumber \\
 \frac{d \lambda_S}{d \ln \mu^2}&=&\frac{1}{16 \pi^2} \left (9 \lambda_S^2+\lambda_{SH}^2 +12y_M^2\lambda_S-12y_M^4\right) \, , \nonumber \\
 \frac{d \lambda_{SH}}{d \ln \mu^2}&=&\frac{1}{16 \pi^2} \left [ \lambda_{SH} \left(2\lambda_{SH}+6 \lambda_H +3 \lambda_S +3 y_t^2  +3 y_T^2+3y_X^2+3y_M^2-\frac{9g_1^2}{20}- \frac{9g_1^2}{20} \right) \right. \nonumber \\
 &-& \left. 6y_T^2y_M^2-6y_X^2y_M^2 \right ] ,
 \end{eqnarray}
\subsubsection{Additional non SM-like quark doublet ${\cal D}_Y$ $(B,Y), \,Y=-5/6$}
The relevant RGE for the Yukawa couplings are
\begin{eqnarray}
\label{eq:rge1singletTXfermion}
\frac{dy_t^2}{d \ln \mu^2}&=& \frac{y_t^2}{16 \pi^2}\left (\frac{9y_t^2}{2}+\frac{3y_B^2}{2}+\frac{9y_Y^2}{2}+\frac{y_M^2}{2}-\frac{17g_1^2}{20}-\frac{9g_2^2}{4}- 8g_3^2 \right)\, ,\nonumber\\
\frac{dy_B^2}{d \ln \mu^2}&=& \frac{y_T^2}{16 \pi^2}\left (\frac{3y_t^2}{2}+\frac{9y_B^2}{2}+ \frac{9y_Y^2}{2}+\frac{y_M^2}{2}-\frac{17g_1^2}{20}-\frac{9g_2^2}{4}- 8g_3^2 \right)\, ,\nonumber\\
\frac{dy_Y^2}{d \ln \mu^2}&=& \frac{y_Y^2}{16 \pi^2}\left (\frac{9y_t^2}{2}+ \frac{9y_Y^2}{2}+ \frac{9y_B^2}{2}+\frac{y_M^2}{2}-2g_1^2-\frac{9g_2^2}{4}- 8g_3^2 \right)\, ,\nonumber\\
\frac{dy_M^2}{d \ln \mu^2}&=& \frac{y_M^2}{16 \pi^2}\left (y_B^2+y_Y^2+\frac{9y_M^2}{2}-\frac{8g_1^2}{5}-8g_3^2\right).
\end{eqnarray}
The Higgs sector RGEs, describing the interactions between the two bosons are:
\begin{eqnarray}
\label{eq:rgesingletTXscalar}
\frac{d \lambda_H}{d \ln \mu^2}&=& \frac{1}{16 \pi^2} \left[  \lambda_H \left (12 \lambda_H+6 y_t^2+6 y_B^2+6y_Y^2-\frac{9 g_1^2}{10}-\frac{9 g_2^2}{2} \right)+ 
 \frac{\lambda_{SH}^2}{4}-3 y_t^4 -3 y_B^4-3y_Y^2 \right. \nonumber \\
 &+& \left. \frac{ 27g_1^4}{400}+\frac{ 9g_2^4}{16} +\frac{ 9g_1^2 g_2^2 }{40}  \right ] \, , \nonumber \\
 \frac{d \lambda_S}{d \ln \mu^2}&=&\frac{1}{16 \pi^2} \left (9 \lambda_S^2+\lambda_{SH}^2 +12y_M^2\lambda_S-12y_M^4\right) \, , \nonumber \\
 \frac{d \lambda_{SH}}{d \ln \mu^2}&=&\frac{1}{16 \pi^2} \left [ \lambda_{SH} \left(2\lambda_{SH}+6 \lambda_H +3 \lambda_S +3 y_t^2  +3 y_B^2+3y_Y^2+6y_M^2-\frac{9g_1^2}{20}- \frac{9g_2^2}{4} \right) \right. \nonumber \\
 &-& \left. 12y_B^2y_M^2-12y_Y^2y_M^2 \right ] ,
 \end{eqnarray}
The coupling constants gain additional terms due to the new fermion in all doublet models as follows:
\begin{eqnarray}
\frac{dg_1^2}{d \ln \mu^2}&=&\frac{g_1^4}{16 \pi^2}\left(\frac{41}{10}+\frac{10}{3}\right )\, , \nonumber \\
\frac{dg_2^2}{d \ln \mu^2}&=&\frac{g_2^4}{16 \pi^2}\left( -\frac{19}{6} +\frac{13}{6} \right ) \, , \nonumber \\
\frac{dg_3^2}{d \ln \mu^2}&=&\frac{g_3^4}{16 \pi^2}\left(-7+\frac{4}{3}\right )\, .
\end{eqnarray}

\subsubsection{Triplet ${\cal T}_X$ $(X, T, B), \,Y=2/3$}
The relevant RGE for the Yukawa couplings are
\begin{eqnarray}
\label{eq:rge1XTBfermion}
\frac{dy_t^2}{d \ln \mu^2}&=& \frac{y_t^2}{16 \pi^2}\left (\frac{9y_t^2}{2}+\frac{9y_T^2}{2}+\frac{9y_X^2}{2}+\frac{3y_B^2}{2}+\frac{y_M^2}{4}-\frac{17g_1^2}{20}-\frac{9g_2^2}{4}- 8g_3^2 \right)\, ,\nonumber\\
\frac{dy_T^2}{d \ln \mu^2}&=& \frac{y_T^2}{16 \pi^2}\left (\frac{9y_t^2}{2}+\frac{9y_X^2}{2}+ \frac{9y_T^2}{2}+\frac{3y_B^2}{2}+\frac{y_M^2}{4}-\frac{17g_1^2}{20}-\frac{9g_2^2}{4}- 8g_3^2 \right)\, ,\nonumber\\
\frac{dy_X^2}{d \ln \mu^2}&=& \frac{y_T^2}{16 \pi^2}\left (\frac{9y_t^2}{2}+ \frac{9y_X^2}{2}+ \frac{9y_T^2}{2}+\frac{3y_B^2}{2}+\frac{y_M^2}{4}-\frac{41g_1^2}{20}-\frac{9g_2^2}{4}- 8g_3^2 \right)\, ,\nonumber\\
\frac{dy_B^2}{d \ln \mu^2}&=& \frac{y_B^2}{16 \pi^2}\left (\frac{3y_t^2}{2}+\frac{9y_B^2}{2}+ \frac{9y_X^2}{2}+ \frac{9y_T^2}{2}+\frac{y_M^2}{4}-\frac{17g_1^2}{20}-\frac{9g_2^2}{4}- 8g_3^2 \right)\, ,\nonumber\\
\frac{dy_M^2}{d \ln \mu^2}&=& \frac{y_M^2}{16 \pi^2}\left (y_T^2+y_X^2+y_B^2+\frac{27y_M^2}{2}-\frac{8g_1^2}{5}-\frac{9g_2^2}{4}-8g_3^2\right).
\end{eqnarray}
The Higgs sector RGEs, describing the interactions between the two bosons are:
\begin{eqnarray}
\label{eq:rgesingletTXscalar}
\frac{d \lambda_H}{d \ln \mu^2}&=& \frac{1}{16 \pi^2} \left[  \lambda_H \left (12 \lambda_H+6 y_t^2+6 y_T^2+6y_X^2-\frac{9 g_1^2}{5}-9 g_2^2 \right)+ 
 \frac{\lambda_{SH}^2}{4}-3 y_t^4 -3 y_T^4-3y_X^2 \right. \nonumber \\
 &-&6y_t^2 y_T^2+ \left. \frac{ 27g_1^4}{200}+\frac{ 9g_2^4}{8} +\frac{ 18g_1^2 g_2^2 }{40}  \right ] \, , \nonumber \\
 \frac{d \lambda_S}{d \ln \mu^2}&=&\frac{1}{16 \pi^2} \left (9 \lambda_S^2+\lambda_{SH}^2 +12y_M^2\lambda_S-12y_M^4\right) \, , \nonumber \\
 \frac{d \lambda_{SH}}{d \ln \mu^2}&=&\frac{1}{16 \pi^2} \left [ \lambda_{SH} \left(2\lambda_{SH}+6 \lambda_H +3 \lambda_S +6 y_t^2  +6 y_T^2+6y_X^2+6y_M^2-\frac{9g_1^2}{10}- \frac{9g_2^2}{2} \right) \right. \nonumber \\
 &-& \left. 6y_T^2y_M^2-6y_B^2y_M^2-6y_X^2y_M^2 \right ] ,
 \end{eqnarray}
\subsubsection{Triplet ${\cal T}_Y$ $( T, B, Y), \, Y=-1/3$}
The relevant RGE for the Yukawa couplings are
\begin{eqnarray}
\label{eq:rge1TBYfermion}
\frac{dy_t^2}{d \ln \mu^2}&=& \frac{y_t^2}{16 \pi^2}\left (\frac{3y_B^2}{2}+\frac{9y_t^2}{2}+\frac{9y_T^2}{2}+\frac{9y_Y^2}{2}+\frac{y_M^2}{2}-\frac{17g_1^2}{20}-\frac{9g_2^2}{4}- 8g_3^2 \right)\, ,\nonumber\\
\frac{dy_T^2}{d \ln \mu^2}&=& \frac{y_T^2}{16 \pi^2}\left (\frac{3y_B^2}{2}+\frac{9y_2^2}{2}+\frac{9y_Y^2}{2}+ \frac{9y_T^2}{2}+\frac{y_M^2}{2}-\frac{17g_1^2}{20}-\frac{9g_2^2}{4}- 8g_3^2 \right)\, ,\nonumber\\
\frac{dy_B^2}{d \ln \mu^2}&=& \frac{y_B^2}{16 \pi^2}\left (\frac{3y_t^2}{2}+\frac{3y_T^2}{2}+\frac{9y_B^2}{2}+ \frac{9y_Y^2}{2}+\frac{y_M^2}{4}-\frac{17g_1^2}{20}-\frac{9g_2^2}{4}- 8g_3^2 \right)\, ,\nonumber\\
\frac{dy_Y^2}{d \ln \mu^2}&=& \frac{y_Y^2}{16 \pi^2}\left (\frac{3y_t^2}{2}+\frac{9y_B^2}{2}+ \frac{9y_Y^2}{2}+ \frac{9y_T^2}{2}+\frac{y_M^2}{2}-\frac{17g_1^2}{20}-\frac{9g_2^2}{4}- 8g_3^2 \right)\, ,\nonumber\\
\frac{dy_M^2}{d \ln \mu^2}&=& \frac{y_M^2}{16 \pi^2}\left (y_T^2+y_B^2+y_Y^2+\frac{15y_M^2}{2}-\frac{2g_1^2}{5}-8g_3^2\right).
\end{eqnarray}
The Higgs sector RGEs, describing the interactions between the two bosons are:
\begin{eqnarray}
\label{eq:rgesingletTXscalar}
\frac{d \lambda_H}{d \ln \mu^2}&=& \frac{1}{16 \pi^2} \left[  \lambda_H \left (12 \lambda_H+6y_t^2+6 y_B^2+6 y_T^2+6y_Y^2-\frac{9 g_1^2}{5}-\frac{9 g_2^2}{2} \right)+ 
 \frac{\lambda_{SH}^2}{4}-6 y_t^4 \right. \nonumber \\
 &-&6 y_T^4-6y_Y^4-3y_B^2-6y_t^2 y_T^2+ \left. \frac{ 27g_1^4}{200}+\frac{ 9g_2^4}{8} +\frac{ 18g_1^2 g_2^2 }{40}  \right ] \, , \nonumber \\
 \frac{d \lambda_S}{d \ln \mu^2}&=&\frac{1}{16 \pi^2} \left (9 \lambda_S^2+\lambda_{SH}^2 +12 y_M^2\lambda_S-12y_M^4\right) \, , \nonumber \\
 \frac{d \lambda_{SH}}{d \ln \mu^2}&=&\frac{1}{16 \pi^2} \left [ \lambda_{SH} \left(2\lambda_{SH}+6 \lambda_H +3 \lambda_S +6 y_t^2  +6 y_T^2+6y_B^2+6y_Y^2+6y_M^2-\frac{9g_1^2}{10}- \frac{9g_2^2}{2} \right) \right. \nonumber \\
 &-& \left. 6y_T^2y_M^2-6y_Y^2y_M^2-6y_B^2y_M^2 \right ] ,
\end{eqnarray}
The coupling constants gain additional terms due to the new fermions in both triplet models  as follows:
\begin{eqnarray}
\frac{dg_1^2}{d \ln \mu^2}&=&\frac{g_1^4}{16 \pi^2}\left(\frac{41}{10}+\frac{16}{5}\right )\, , \nonumber \\
\frac{dg_2^2}{d \ln \mu^2}&=&\frac{g_2^4}{16 \pi^2}\left( -\frac{19}{6} +4\right ) \, , \nonumber \\
\frac{dg_3^2}{d \ln \mu^2}&=&\frac{g_3^4}{16 \pi^2}\left(-7+2\right )\, .
\end{eqnarray}

\newpage
\subsection{Oblique Parameters $\mathbb{S}$ and $\mathbb{T}$ in the Scalar Sector}
We give below the conventional shifts in oblique parameters $\mathbb{S}$ and $\mathbb{T}$ from the SM. The original expressions appeared in \cite{Xiao:2014kba} with only one singlet representation. Additional contributions to two point functions are also included in \cite{Kanemura:2015fra} for further analysis.

The explicit expressions for the $\Delta \mathbb{S}$ and $\Delta \mathbb{T}$  parameters for the SHM, including the extra singlet scalar representation, but without the vector-like quarks, are
\label{eq:Tscalar}
\begin{equation}
\Delta \mathbb{T}=\mathbb{T}_{SH}-\mathbb{T}_{SM}=-\frac{3s_{\phi}^{2}}{16\pi c_{w}^{2}}\left([t_{S}]-[t_{H}] \right)
\end{equation}
where 
\begin{equation*}
[t_{S}]=\left((m_{S}^{2}-m_{Z}^{2})(m_{S}^{2}-m_{W}^{2}) \right)^{-1}\left[m_{S}^{4}\ln(m_{S}^{2})-\frac{(m_{S}^{2}-m_{W}^{2})m_{Z}^{2}\ln(m_{Z}^{2})-m_{W}^{2}\ln(m_{W}^{2})c_{w}^{2}(m_{S}^{2}-m_{Z}^{2})}{s_{w}^{2}} \right]
\end{equation*}
and similarly for $[t_{H}]$ function, with the replacement $m_{S} \to m_{H}$.

\label{eq:Sscalar}
\begin{equation}
\Delta \mathbb{S}=\mathbb{S}_{SH}-\mathbb{S}_{SM}=\frac{s_{\phi}^{2}}{12\pi}\left(2\ln(\frac{m_{S}}{m_{H}})+[s_{S}]-[s_{H}] \right)
\end{equation}
where 
\begin{equation*}
[s_{S}]=\frac{m_{Z}^{4}(9m_{S}^{2}+m_{Z}^{2})}{(m_{S}^{2}-m_{Z}^{2})^{3}}\ln(\frac{m_{S}^{2}}{m_{Z}^{2}})-\frac{m_{Z}^{2}(4m_{S}^{2}+6m_{Z}^{2})}{(m_{S}^{2}-m_{Z}^{2})^{2}}
\end{equation*}
and similarly for $[s_{H}]$ function, with the replacement $m_{S} \to m_{H}$.

\subsection{Vector-Like Quark contributions to the $\mathbb{S}$ and $\mathbb{T}$ parameters}
\label{sec:STapex}
\subsubsection{Singlet ${\cal U}_1~(T)$, $Y=2/3$}
\label{eq:STsingletU}
\begin{eqnarray}
\Delta \mathbb{T}&=&\frac{m_{t}^{2}N_{c}(s_{L}^{t})^{2}}{16\pi c_{w}^{2}s_{w}^{2}m_{Z}^{2}}\left[(x_{T}^{2}(s_{L}^{t})^{2} - (c_{L}^{t})^{2}-1+4(c_{L}^{t})^{2}\frac{m_{T}^{2}}{m_{T}^{2}-m_{t}^{2}}\ln(x_{T}) \right], \nonumber \\
\Delta \mathbb{S}&=&\frac{N_{c}(s_{L}^{t})^{2}}{18\pi}\left(\frac{(c_{L}^{t})^{2}}{(x_{T}-1)^{3}} \left[2\ln(x_{T})(3-9x_{T}^{2}-9x_{T}^{4}+3x_{T}^{6}) +5-27x_{T}^{2}-27x_{T}^{4}-5x_{T}^{6}\right]  \right. \nonumber \\
 &-& \left. 2\ln(x_{T}) \right)
\end{eqnarray}

\subsubsection{Singlet ${\cal D}_1~(B)$, $Y=-1/3$}
\label{eq:STsingletD}
\begin{eqnarray}
\Delta \mathbb{T}&=&\frac{m_{t}^{2}N_{c}x_{B}}{16\pi c_{w}^{2}s_{w}^{2}m_{Z}^{2}(x_{B}-1)}\left[(s_{L}^{b})^{4}(x_{B}-1) -2(s_{L}^{b})^{2}\ln(x_{B})  \right], \nonumber \\
\Delta \mathbb{S}&=&\frac{N_{c}(s_{L}^{b})^{2}}{18\pi}\left[2\ln(\frac{m_{b}}{m_{B}})(3(s_{L}^{b})^{2}-4)-5(c_{L}^{b})^{2} \right]
\label{eq:TSD1}
\end{eqnarray}

\subsubsection{Doublet ${\cal D}_2$ $(T,B), \,Y=1/6$}
\label{eq:STdoubletTB}
\begin{eqnarray}
\Delta \mathbb{T}&\simeq&\frac{m_{t}^{2}N_{c}(s_{R}^{t})^{2}}{8\pi c_{w}^{2}s_{w}^{2}m_{Z}^{2}}\left[2\ln(x_{T})-2 \right], \nonumber \\
\Delta \mathbb{S}&\simeq&\frac{N_{c}}{18\pi}\left[(s_{R}^{b})^{2}\left(2\ln(x_{T})-2\ln(x_{b})-2 \right) +(s_{R}^{t})^{2}\left(4\ln(x_{T})-7 \right) \right]
\end{eqnarray}

\subsubsection{Doublet ${\cal D}_X$ $(X,T), \,Y=7/6$}
\label{eq:STdoubletXT}
\begin{eqnarray}
\Delta \mathbb{T}&\simeq&\frac{m_{t}^{2}N_{c}(s_{R}^{t})^{2}}{8\pi c_{w}^{2}s_{w}^{2}m_{Z}^{2}(x_{T}-1)}\left[\ln\left((c_{R}^{t})^{2}+\frac{(s_{R}^{t})^{2}}{x_{T}}\right) -\ln(x_{T})[x_{T}+\mathcal{O}(x_{T}^{-4})]\right], \nonumber \\
\Delta \mathbb{S}&\simeq&\frac{N_{c}(s_{R}^{t})^{2}}{18\pi}\left[ 3+\ln(x_{T})+\mathcal{O}\left(\frac{(s_{R}^{t})^{4}}{x_{B}}\right)\right]
\label{eq:TSDX}
\end{eqnarray}

\subsubsection{Doublet ${\cal D}_Y$ $(B,Y), \,Y=-5/6$}
\label{eq:STdoubletBY}
\begin{eqnarray}
\Delta \mathbb{T}&\simeq&\frac{m_{t}^{2}N_{c}x_{B}}{128\pi c_{w}^{2}s_{w}^{2}m_{Z}^{2}}\left[-16c_{R}^{b}\left(-3+c_{R}^{2b}\cot_{R}^{b}\ln(c_{R}^{b}) \right) + s_{R}^{b}\left(-13-20c_{R}^{2b}+4c_{R}^{2b} \right) \right], \nonumber \\
\Delta S&\simeq&\frac{N_{c}}{144\pi}\left[-3\ln(x_{B})+20\ln(c_{R}^{b})+\ln(x_{b})+19+c_{R}^{4b}\left(5+3\ln(x_{b}) \right)+4c_{R}^{2b}\left(-6-\ln(x_{b}) \right) \right]
\end{eqnarray}

\subsubsection{Triplet ${\cal T}_X$ $(X, T, B), \,Y=2/3$}
\label{eq:STtripletXTB}
\begin{eqnarray}
\Delta  \mathbb{T}&\simeq&\frac{m_{t}^{2}N_{c}(s_{L}^{t})^{2}}{16\pi c_{w}^{2}s_{w}^{2}m_{Z}^{2}}\left[6\ln(x_{T})-10  +\mathcal{O}((s_{L}^{t})^{4},(c_{L}^{t})^{4},x_{T}^{-4})\right], \nonumber \\
\Delta S&\simeq&-\frac{N_{c}(s_{L}^{t})^{2}}{18\pi}\left[ 9-6\ln(x_{T})+4\ln(x_{b}) +\mathcal{O}((c_{L}^{t})^{4},(c_{L}^{t})^{2}(s_{L}^{t})^{2})\right]
\end{eqnarray}

\subsubsection{Triplet ${\cal T}_Y$ $( T, B, Y), \, Y=-1/3$}
\label{eq:STtripletTBY}
\begin{eqnarray}
\Delta \mathbb{T}&\simeq&-\frac{m_{t}^{2}N_{c}(s_{L}^{t})^{2}}{16\pi c_{w}^{2}s_{w}^{2}m_{Z}^{2}}\left[2\ln(x_{T})-6  +\mathcal{O}((s_{L}^{t})^{4},(c_{L}^{t})^{2}(s_{L}^{t})^{2},x_{T}^{-3})\right], \nonumber \\
\Delta \mathbb{S}&\simeq&\frac{N_{c}(s_{L}^{t})^{2}}{18\pi}\left[2\ln(x_{T})+4 +\mathcal{O}((c_{L}^{t})^{4},(s_{L}^{t})^{4},(c_{L}^{t})^{2}(s_{L}^{t})^{2})\right]
\end{eqnarray}
where $x_{i}=\frac{m_F}{m_t}$ for all representations.



\newpage

\begin{thebibliography}{}\label{sec:bibs}
\bibitem{cms-atlas}
 G.~Aad {\it et al.}  [ATLAS Collaboration],
  Phys.\ Lett.\ B {\bf 716} 1 (2012);
  S.~Chatrchyan {\it et al.}  [CMS Collaboration],
  Phys.\ Lett.\ B {\bf 716} 30 (2012).

\bibitem{stability}
G. Degrassi  {\it et al.}, 
JHEP 1208 (2012), 098; 
D.  Buttazzo {\it et al.}, 
JHEP 1312 , 089 (2013). 

\bibitem{Gunion:2002zf} 
  J.~F.~Gunion and H.~E.~Haber,
  Phys.\ Rev.\ D {\bf 67}, 075019 (2003)
  
\bibitem{inflation}
I. Masina and A. Notari A., 
Phys. Rev. D {\bf 85} 123506 (2012);  
Isidori G. {\it et al.}, 
Phys. Rev. D {\bf 77}  025034 (2008); 
F. L.  Bezrukov and M. Shaposhnikov,  
Phys. Lett. B 659, 703 (2008);
K. Kamada  {\it et al.} , 
Phys. Rev. D {\bf 86}, 023504 (2012). 


\bibitem{AguilarSaavedra:2002kr} 
  J.~A.~Aguilar-Saavedra,
  Phys.\ Rev.\ D {\bf 67}, 035003 (2003)
  Erratum: [Phys.\ Rev.\ D {\bf 69}, 099901 (2004)].

\bibitem{Aguilar-Saavedra:2013qpa} 
  J.~A.~Aguilar-Saavedra, R.~Benbrik, S.~Heinemeyer and M.~Perez-Victoria,
  Phys.\ Rev.\ D {\bf 88}, 094010 (2013);
  J.~A.~Aguilar-Saavedra,
  EPJ Web Conf.\  {\bf 60}, 16012 (2013).


\bibitem{Ellis:2014dza}
S.~ A.~R. Ellis, R.~M. Godbole, S. Gopalakrishna, and J.~D.
  Wells,
 JHEP, 09:130, (2014); S.~Dawson and E.~Furlan.
 Phys. Rev., D {\bf 86} 015021, 2012; J.~A. Aguilar-Saavedra, R.~Benbrik, S.~Heinemeyer, and M.~Perez-Victoria,
 Phys. Rev., D {\bf 88} 094010  (2013); G. Cacciapaglia, A. Deandrea, D. Harada, and Y. Okada,
JHEP, 11:159  (2010).


\bibitem{Gopalakrishna:2013hua}
S.~Gopalakrishna, T.~Mandal, S.~Mitra and G.~Moreau,
JHEP \textbf{08} (2014), 079
doi:10.1007/JHEP08(2014)079
[arXiv:1306.2656 [hep-ph]].


\bibitem{Contino:2006qr}
  R.~Contino, L.~Da Rold and A.~Pomarol,
  Phys.\ Rev.\ D {\bf 75}  055014 (2007); 
  B.~A.~Dobrescu and C.~T.~Hill,
  Phys.\ Rev.\ Lett.\  {\bf 81}  2634 (1998).

\bibitem{Anastasiou:2009rv}
  C.~Anastasiou, E.~Furlan and J.~Santiago,
  Phys.\ Rev.\ D {\bf 79} 075003 (2009). 
  
    \bibitem{ArkaniHamed:2002qy}
N.~Arkani-Hamed, A.~G. Cohen, E.~Katz, and A.~E. Nelson,
 JHEP, 07:034   (2002);
Jay Hubisz and Patrick Meade.
 Phys. Rev. D {\bf 71} 035016  (2005).


\bibitem{Martin:2009bg}
  S.~P.~Martin,
  Phys.\ Rev.\ D {\bf 81} (2010) 035004; 
  S.~P.~Martin,
  Phys.\ Rev.\ D {\bf 82} (2010) 055019.


\bibitem{Couture:2017mbd} 
  G.~Couture, M.~Frank, C.~Hamzaoui and M.~Toharia,
  Phys.\ Rev.\ D {\bf 95}, no. 9, 095038 (2017);
  A. Angelescu, A. Djouadi, and G. Moreau.
 Eur. Phys. J. {\bf C} 76(2): 99, 2016.

  
\bibitem{Xiao:2014kba} 
  M.~L.~Xiao and J.~H.~Yu,
  Phys.\ Rev.\ D {\bf 90}, no. 1, 014007 (2014); 
  Addendum: [Phys.\ Rev.\ D {\bf 90}, no. 1, 019901 (2014)].
  
\bibitem{Dhuria:2015ufo} 
  M.~Dhuria and G.~Goswami,
  Phys.\ Rev.\ D {\bf 94}, no. 5, 055009 (2016).

\bibitem{Zhang:2015uuo} 
  J.~Zhang and S.~Zhou,
  Chin.\ Phys.\ C {\bf 40}, no. 8, 081001 (2016).
  
\bibitem{Carmi:2012yp} 
  D.~Carmi, A.~Falkowski, E.~Kuflik and T.~Volansky,
  JHEP {\bf 1207}, 136 (2012);
  S.~Dawson and E.~Furlan,
  Phys.\ Rev.\ D {\bf 86}, 015021 (2012);
  J.~Kang, P.~Langacker and B.~D.~Nelson,
  Phys.\ Rev.\ D {\bf 77}, 035003 (2008);
  S.~Fajfer, A.~Greljo, J.~F.~Kamenik and I.~Mustac,
  JHEP {\bf 1307}, 155 (2013).
  
\bibitem{Altarelli:1994rb} 
  G.~Altarelli and G.~Isidori,
  Phys.\ Lett.\ B {\bf 337}, 141 (1994).
  
\bibitem{Tang:2013bz} 
  Y.~Tang,
  Mod.\ Phys.\ Lett.\ A {\bf 28}, 1330002 (2013).
  
\bibitem{Sirlin:1985ux} 
  A.~Sirlin and R.~Zucchini,
  Nucl.\ Phys.\ B {\bf 266}, 389 (1986).
  
\bibitem{Hempfling:1994ar} 
  R.~Hempfling and B.~A.~Kniehl,
  Phys.\ Rev.\ D {\bf 51}, 1386 (1995).
  
\bibitem{Chen:2017hak}
C.~Y.~Chen, S.~Dawson and E.~Furlan,
Phys. Rev. D \textbf{96} (2017) no.1, 015006.

 
\bibitem{ATLAS-CONF-2016-101}
  A.~M.~Sirunyan {\it et al.} [CMS Collaboration],
  JHEP {\bf 1711}, 085 (2017);
  A.~M.~Sirunyan {\it et al.} [CMS Collaboration],
  JHEP {\bf 1705}, 029 (2017); 
  V.~Khachatryan {\it et al.} [CMS Collaboration],
  Phys.\ Rev.\ D {\bf 93}, no. 11, 112009 (2016)
  G.~Aad {\it et al.} [ATLAS Collaboration],
  JHEP {\bf 1602}, 110 (2016);
   G.~Aad {\it et al.}  [ATLAS Collaboration],
Phys. Lett. B {\bf 758} 249 (2016);
  G.~Aad {\it et al.}  [ATLAS Collaboration],
 Eur. Phys. J. C {\bf 76} (8) 442  (2016);
  G.~Aad {\it et al.}  [ATLAS Collaboration],
\ JHEP, 02:110  (2015);
  G.~Aad {\it et al.}  [ATLAS Collaboration],
 JHEP, 08:105  (2015);
  G.~Aad {\it et al.}  [ATLAS Collaboration],
Phys. Rev. D {\bf  91}(11):112011  (2015);
V. Khachatryan {\it et~al.} [CMS Collaboration], 
 Phys. Rev. D {\bf 93}(11) 112009  (2016);
V. Khachatryan {\it et~al.} [CMS Collaboration],
 JHEP, 06 080  (2015);
S. Chatrchyan {\it et~al.} [CMS Collaboration],
 Phys. Rev. Lett., 112(17) 171801  (2014);
S. Chatrchyan {\it et~al.} [CMS Collaboration],
 Phys. Lett. B {\bf  729} 149  (2014);
  G.~Aad {\it et al.}  [ATLAS Collaboration],
Technical Report ATLAS-CONF-2017-015 (2017).

\bibitem{moriond2018}
N.~ Nikiforou, 
talk, Rencontres de Moriond (2018).

\bibitem{ATLAS:2017vdo}
M.~Aaboud \textit{et al.} [ATLAS],
JHEP \textbf{08} (2017), 052.


\bibitem{RGEpt1}
M.~E. Machacek and M.~T. Vaughn,
 Nucl. Phys. B {\bf 222} (1) 83 (1983);
M.~E. Machacek and M.~T. Vaughn,
Nucl. Phys. B {\bf 236} (1) 221 (1984);
M.~E. Machacek and M.~T. Vaughn,
 Nucl. Phys. B {\bf 249}(1) 70 (1985).

\bibitem{Lavoura:1992np}
L.~Lavoura and J.~P.~Silva,
Phys. Rev. D \textbf{47} (1993), 2046-2057.


\bibitem{Hagiwara1995}

K.Hagiwara, S.Matsumoto, D.Haidt and C. S. Kim, Z. Phys. C{\bf64}, 559(1994)[Erratum-ibid.
 C {\bf68}, 352 (1995).
  
\bibitem{Kanemura:2015fra}
S.~Kanemura, M.~Kikuchi and K.~Yagyu,
Nucl. Phys. B \textbf{907} (2016), 286-322.


\bibitem{Bahrami:2014ska}
S.~Bahrami and M.~Frank,
Phys. Rev. D \textbf{90} (2014) no.3, 035017.

\end{thebibliography}
\end{document}